\documentclass[twocolumn]{aa}  

\usepackage{graphicx}
\usepackage{txfonts}
\usepackage[mathlines,switch]{lineno}
\begin{document}

   \title{Surveys of clumps, cores, and condensations in Cygnus-X:}

   \subtitle{Searching for Keplerian disks on the scale of 500 au}

   \author{Xing Pan\inst{1,2,3}, Keping Qiu\inst{1,2}, Qizhou Zhang\inst{3}}

   \institute{School of Astronomy and Space Science, Nanjing University, 163 Xianlin Avenue, Nanjing 210023, P.R.China \\ \email{kpqiu@nju.edu.cn} \and 
   Key Laboratory of Modern Astronomy and Astrophysics (Nanjing University), Ministry of Education, Nanjing 210023, P.R.China \and
   Center for Astrophysics $\vert$ Harvard \& Smithsonian, 60 Garden Street, Cambridge, MA, 02138, USA}
             
   \date{}

 
  \abstract
   {Over the past decades, observational evidence of circumstellar disks around massive protostars has been steadily accumulating. However, there have also been cases of non-detections in high-mass star-forming regions, leaving the role and prevalence of disks around massive protostars still uncertain.}
   {The aim of this work is to investigate the substructures of the previously identified 2000-au-scale rotating structures around massive protostars and search for the embedded Keplerian disk inside.}
   {We used high-resolution ($\sim0.2\arcsec$) NOrthern Extended Millimeter Array (NOEMA) observations to study the 1.3 mm continuum and molecular line emission of five massive dense cores in the Cygnus-X cloud complex. Four cores host 2000-au-scale rotating structures previously identified as disk candidates in lower-resolution SMA observations, while the remaining core with no evidence for a disk serves as a comparison.}
   {With a resolution of 300 au, the 1.3 mm continuum emission reveals varying levels of fragmentation in our sample, with fragment radii ranging from 150 to 800 AU. The emission of the CO J=2-1 transition shows that 13 fragments are associated with uni- or bipolar outflows, but only seven are detected in the $\mathrm{CH_3CN}$ emission. We find velocity gradients across two fragments perpendicular to the outflow axis and their position-velocity (PV) diagrams along the velocity gradient resemble the Keplerian rotation. Fitting the velocity profiles in the PV diagrams with a Keplerian model, we obtain protostellar masses for the two disks. Both disks have gas masses lower than 1/3 of the protostellar masses and Toomre Q values are higher than 1, indicating that the disks are globally stable. Among the other sources detected in the $\mathrm{CH_3CN}$ emission, some show velocity fields indicative of gas flows connecting multiple systems or outflowing gas, while others show no clear velocity gradient.}
   {In this work, we confirm the existence of two small, stable disks in Keplerian-like rotation at scales of 500 au out of four previously identified disk candidates from the SMA observations at coarser resolution. The lack of evidence for Keplerian disks in other disk candidates identified from the SMA data suggests that rotational signatures observed at 2000 au scales do not necessarily imply the presence of Keplerian disks at smaller scales. Therefore, higher-resolution and higher-sensitivity observations are essential to definitively identify Keplerian disks on smaller scales.}

   \keywords{}

   \maketitle
%
\section{Introduction}\label{sec:intro}
Disk-mediated accretion has been widely regarded as a common mechanism for star formation across different stellar masses. A collapsing cloud with initial angular momentum flattens along the dimensions perpendicular to the rotation axis, which leads to the formation of a disk \citep{Shu1987}. Numerical simulations \citep[e.g.,][]{Kuiper2011,Klassen2016,Rosen2016,Harries2017,Meyer2018,Kuiper2018}{}{} predict that accretion disks channel material to the central protostar, and help overcome radiation pressure that could otherwise halt accretion. This process is particularly important in high-mass star formation ($M_*>8M_\odot$), where protostars may continue the mass growth after the onset of hydrogen burning in their interior.

Numerous disks around low- ($M_*<2M_\odot$) and intermediate-mass young stars ($2M_\odot<M_*<8M_\odot$) have been detected, typically ranging in size from 10 to 100 AU \citep[e.g.,][]{Maury2019,Andrews2009,Ansdell2016,Tripathi2017,Andrews2018}. However, direct detection of disks around massive protostars remains challenging due to their rarity, greater distances, rapid evolution, and the crowded environments in which they form. Additionally, most OB-type stars are in binary or multiple systems \citep[e.g.,][]{Apai2007,Chini2012,Offner2023}, further complicating the kinematic interpretation of their surrounding gas if the resolution is insufficient to resolve individual components.

Despite observational challenges, growing evidence supports the existence of disks around high-mass protostars \citep[e.g.,][]{Zapata2010,Sanchez-Monge2013,Cesaroni2014,Beltran2014,Johnston2015,Ilee2016,Girart2017,Cesaroni2017,Sanna2019,Lu2022,Pan2024}. These disks typically have radii of several thousand au, though they are often observed at resolutions of 1000 au or coarser \citep[see Table 2 in][]{Beltran2016}. One of the most compelling examples is IRAS 20126+4104, which has been  studied extensively \citep{Zhang1998,Cesaroni1999,Qiu2008,Keto2010,Cesaroni2014,Chen2016} and confirmed to host a large ($\sim$1000 au) Keplerian disk rotating around a $\sim$$10^4 L_\odot$ protostar with observations at different resolutions from 600 au to 5000 au. On the other hand, recent ALMA observations with resolutions of $\lesssim$100 au have revealed Keplerian disks around 100 au in size, such as Orion Source I \citep{Ginsburg2018}, MonR2-IRS2 \citep{Jimenez-Serra2020}, and G17.64+0.16 \citep{Maud2018}. Meanwhile, some studies at similar resolutions have found no evidence for disks in high-mass star-forming regions \citep{Ginsburg2017,Cesaroni2017,Goddi2020}. These observational studies have shown a diverse behavior of circumstellar gas near different high-mass protostars that still needs to be investigated. 

Theoretical studies on accretion mechanism in massive star formation, incorporating various physical processes and initial conditions, also show diverse disk configurations. \cite{Oliva2020} presented a self-gravity-radiation-hydrodynamic simulation showing the formation of a massive protostar surrounded by a fragmenting Keplerian-like accretion disk with spiral arms. \cite{Kuiper2018} conducted the first hydrodynamic simulations of high-mass star formation, incorporating radiative feedback from both radiation forces and photoionization, and confirmed the existence of large Keplerian disks with radii expanding over time to a few thousand au. When magnetic fields are included, magnetohydrodynamic (MHD) simulations \citep[e.g.,][]{Commercon2011, Seifried2011,Oliva2023} show that strong magnetic fields suppress Keplerian disk formation and fragmentation, whereas weak magnetic fields promote a highly clustered environment. To decipher the accretion mechanism in high-mass star formation, we still need to increase the sample size of disk candidates around massive protostars.

There have been large surveys to identify disk candidates around massive protostars. For instance, the IRAM and NOrthern Extended Millimeter Array (NOEMA) large program CORE \citep{Beuther2018} targeted 20 high-mass star-forming regions at high angular resolutions ($\sim0.4\arcsec$), and identified 13 disk candidates on the scale of 1000 au \citep{Ahmadi2023}. \cite{Pan2024} also established a large sample of 48 massive dense cores (MDCs) in the Cygnus-X molecular cloud using 1.3 mm observations with the Submillimeter Array (SMA) at a resolution of 1.8$\arcsec$ (about 2700 au at a distance of 1.4 kpc for the Cygnus-X cloud). All sources are located within the same molecular cloud complex and were uniformly observed. \cite{Pan2024} identified nine disk candidates out of 27 2000-au-scale condensations with gas masses around ten solar masses. To resolve substructures in these disk-like sources and search for true Keplerian disks, we conducted follow-up 1.3 mm observations using NOEMA at a higher resolution (0.2$\arcsec$, about 280 au at the distance of 1.4 kpc). As part of the Surveys of Clumps, Cores, and Condensations in CygnUS-X (CENSUS) project \citep[PI: Keping Qiu,][]{Cao2019,Wang2022,Zhang2024,Yang2024,Pan2024} project, this paper aims to reveal substructures within 0.01-pc-scale condensations and to search for smaller-scale Keplerian disks using dense gas tracers like $\mathrm{CH_3CN}$. This paper is structured as follows. Section \ref{sec:obs} summarizes the observations and data reduction. Section \ref{sec:results} presents the 1.3 mm continuum and line emission maps. In Section \ref{sec:discussion}, we discusses the detection and stability of Keplerian disks. Summaries are presented in Section \ref{sec:summary}.

\section{Observation} \label{sec:obs}
Our sample was established based on the identification of disk candidates in the Cygnus-X cloud using the SMA observations reported in \cite{Pan2024}. We found nine condensations ($\sim$ 2000 au) showing evidence of rotation within 48 massive dense cores (MDCs, $R\sim10^4$ au, $M_\mathrm{gas}>40~M_\odot$) in the Cygnus-X cloud. Among them, we selected four MDCs (N30, N56, N63, NW14) exhibiting clear rotation evidence to resolve the underlying Keplerian disk (on the order of 100 au) inside them and one MDC (N68) without rotation evidence as a comparison. Table \ref{tab:obs} lists the coordinates of the central positions of these sources.

\begin{figure}[!ht]
    \centering
    \includegraphics[width=1.0\linewidth]{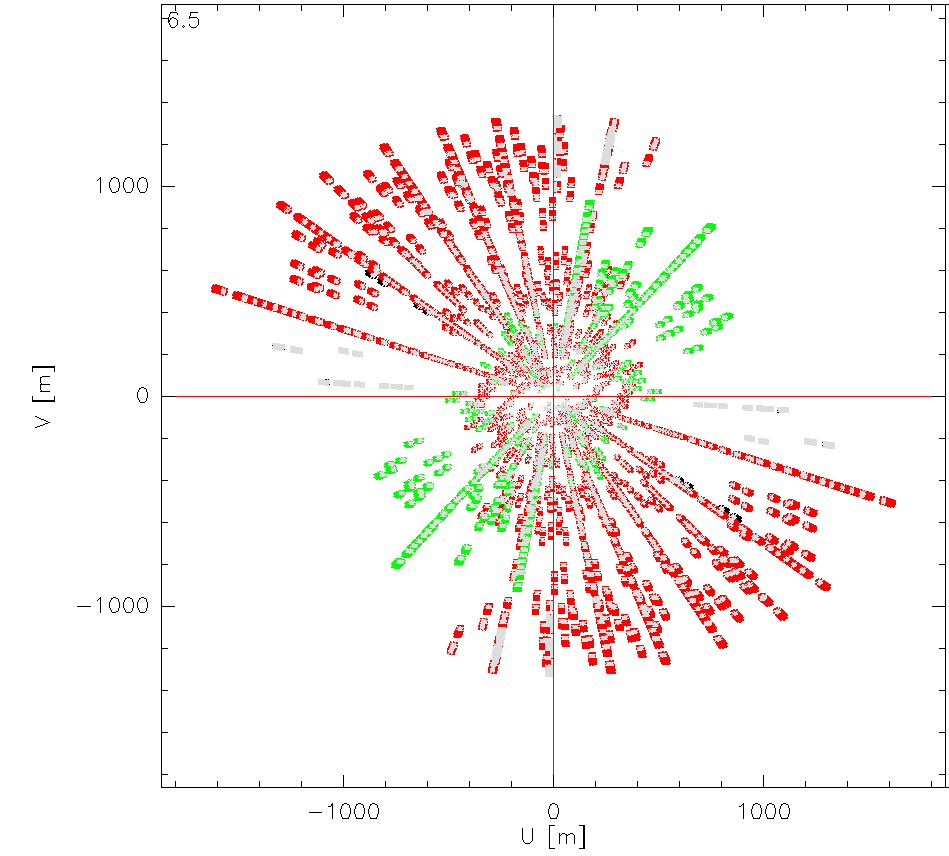}
    \caption{$uv$-coverage for N30. The different colors represent observations from different tracks.}
    \label{fig:N30-uvcov}
\end{figure}

All sources are observed in the tracking-sharing mode between January 2024 and February 2024. To achieve the angular resolution needed to resolve the Keplerian disk, we observed in the A-configuration with baselines ranging from 45 m to 1663 m. The $uv$-coverage for N30 is shown in Fig. \ref{fig:N30-uvcov}. Two quasars, 2013+370 and 2005+403,  were observed for time-dependent phase and amplitude calibration in each track. Strong quasars (e.g., 3C84, 3C279, or 2200+420) were used as bandpass calibrators. MWC349 was observed for absolute flux calibration, assuming a model flux of 1.93 Jy at 1.3 mm. 

\begin{table*}[!ht]
\centering
\caption{List of massive dense cores observed with NOEMA.}
\begin{tabular}{ccccccccc}
\hline
\hline
Source\tablefootmark{a} & Field\tablefootmark{b}  & R.A. & Dec. & Beam\tablefootmark{c} & $\sigma_\mathrm{rms}$ & $I_\mathrm{peak}$\tablefootmark{d}\\
& ID & (J2000) & (J2000) & ($\arcsec$, PA) & (mJy beam$^{-1}$) & (mJy beam$^{-1}$) \\
\hline
N30 & Field 7 & 20:38:36.42 & +42:37:34.62 & $0.21\arcsec\times0.12\arcsec(69^\circ)$ & 0.70 & 113.85 \\
N56 & Field 8 & 20:39:16.74 & +42:16:09.31 & $0.23\arcsec\times0.12\arcsec(70^\circ)$ & 0.28 & 53.78 \\
N63 & Field 9 & 20:40:05.39 & +41:32:13.02 & $0.24\arcsec\times0.12\arcsec(74^\circ)$ & 0.60 & 167.60 \\
N68 & Field 11 & 20:40:33.57 & +41:59:01.07 & $0.26\arcsec\times0.12\arcsec(66^\circ)$ & 0.18 & 12.12 \\
NW14 & Field 16 & 20:24:31.68 & +42:04:22.51 & $0.21\arcsec\times0.12\arcsec(63^\circ)$ & 0.20 & 44.10 \\
\hline
\end{tabular}
\tablefoot{\tablefootmark{a} Massive dense cores identified from \cite{Motte2007}. \tablefootmark{b} Field ID for the SMA observations from \cite{Pan2024}. \tablefootmark{c} The synthesized beam with a robust parameter of 0.5. \tablefootmark{d} The peak intensity of each source in Fig. \ref{fig:1.3mm_cont}}
\label{tab:obs}
\end{table*}

We used the 250 kHz (corresponding to a velocity resolution of $\sim$0.3 $\mathrm{km~s^{-1}}$) correlator mode of PolyFiX at Band 3 (230 GHz). It contains two sidebands: The lower sideband (LSB) and the upper sideband (USB) are centered at 217.756 and 233.241 GHz, respectively, each with a bandwidth of 7.744 GHz. Important lines covered in the bandwidth are summarized in Table \ref{tab:lineinfo}.

\begin{table}[!ht]
    \centering
    \caption{Information on the important lines covered in the bandwidth.}
    \begin{tabular}{cc}
    \hline
    \hline
    Spectral Line & Rest Frequency \\
     & (GHz) \\
    \hline
    $\mathrm{CH_3CN\ (12_0-11_0)}$ & 220.747  \\
    $\mathrm{CH_3CN\ (12_1-11_1)}$ & 220.743  \\
    $\mathrm{CH_3CN\ (12_2-11_2)}$ & 220.730  \\
    $\mathrm{CH_3CN\ (12_3-11_3)}$ & 220.709  \\
    $\mathrm{CH_3CN\ (12_4-11_4)}$ & 220.679  \\
    $\mathrm{CH_3CN\ (12_5-11_5)}$ & 220.641  \\
    $\mathrm{CH_3CN\ (12_6-11_6)}$ & 220.594  \\
    $\mathrm{HNCO~(10_{1,9}-9_{1,8})}$ & 220.585 \\
    $\mathrm{SiO~(5-4)}$ & 217.105 \\
    $\mathrm{C^{18}O~(2-1)}$ & 219.560 \\
    $\mathrm{CH_3OH~(8_{0,6}-7_{1,8})}$ & 220.078 \\
    $\mathrm{CO~(2-1)}$ & 230.538 \\
    \hline
    \end{tabular}
    \label{tab:lineinfo}
\end{table}

Calibration and imaging of the NOEMA observations were conducted using the CLIC and MAPPING program of the Grenoble Image and Line Data Analysis Software (GILDAS\footnote{\url{https://www.iram.fr/IRAMFR/GILDAS/}}). We separated the continuum and spectral lines in the \emph{uv} domain by averaging all the line-free channels to obtain the continuum data and subtracting that continuum level from each spectral line dataset. For a compromise between optimized sensitivity and sidelobe suppressing, the continuum and spectral line images were made with the ROBUST weighting parameter set to 0.5. The final synthesized beam sizes for the continuum images are listed in Table \ref{tab:obs}. The beam sizes for the spectral line images are similar to that for the continuum. We also applied self-calibration in MAPPING to improve the dynamical range of the continuum images. The 1$\sigma$ rms noise and peak intensity of the final continuum images for each source are listed in Table \ref{tab:obs}.

\section{Results}\label{sec:results}
\subsection{Dust continuum}\label{subsec:dustcont}

\begin{figure*}[!htp]
    \centering
    \includegraphics[width=0.95\textwidth]{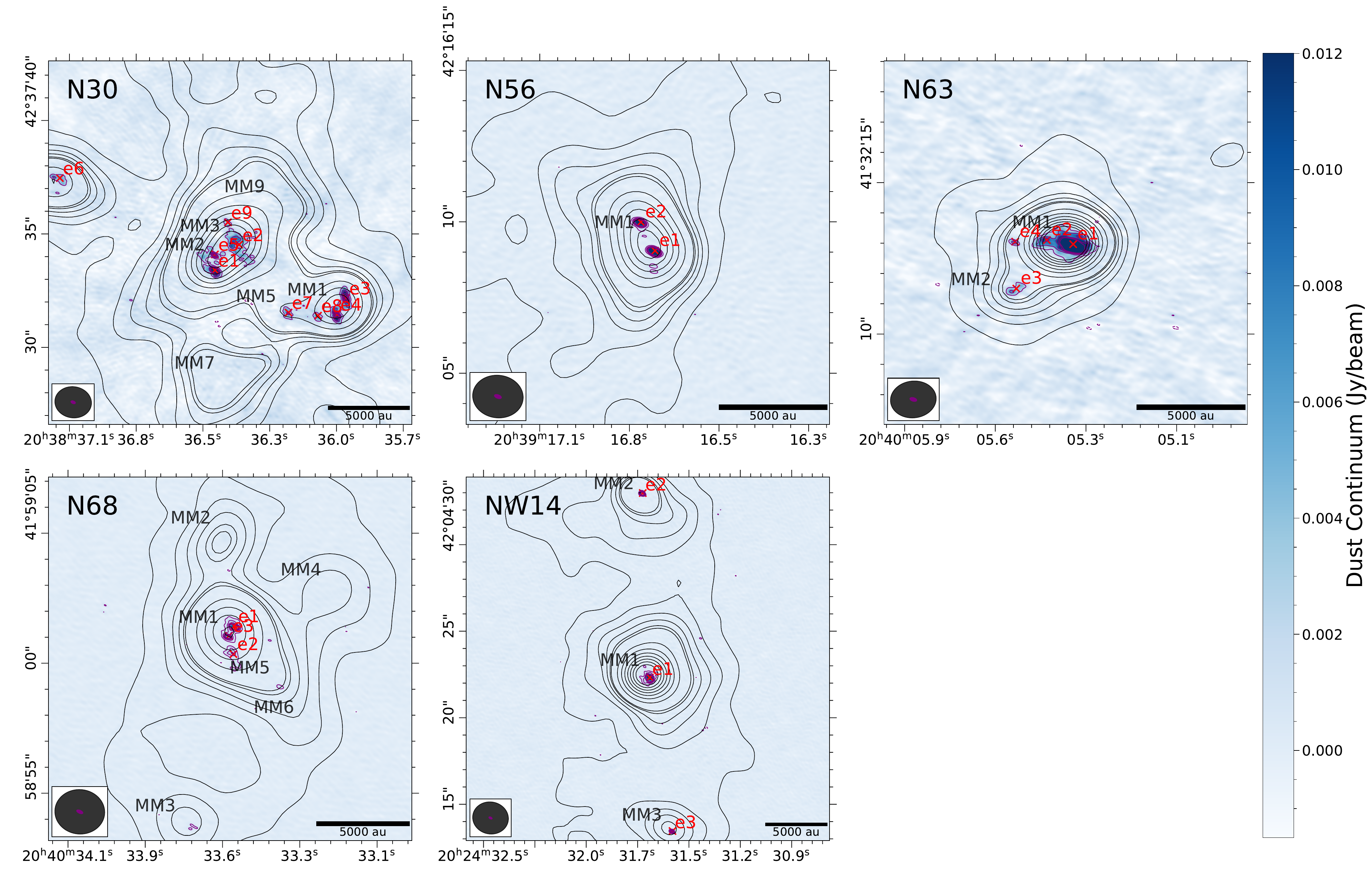}
    \caption{The 1.3 mm continuum images for sources in our sample obtained with NOEMA. The purple contours start from 4$\sigma$ and increase in steps of 6$\sigma$. The red crosses represent the position of fragments identified by Dendrogram. We also overlay the low-resolution ($\sim$1.8$\arcsec$) SMA 1.3 mm continuum emission in black contours. The black labels starting with MM indicate the 2000 au scale condensations identified with the SMA observations in \cite{Pan2024}. The synthesized beam of each image is shown in the bottom left. The purple and black ellipses indicates the beam size of the NOEMA and SMA observations, respectively.}
    \label{fig:1.3mm_cont}
\end{figure*}

Fig. \ref{fig:1.3mm_cont} shows the NOEMA 1.3 mm continuum emission of five massive dense cores (MDCs) in our sample. \citet{2019ApJS..241....1C} identified hundreds of 0.01-pc-scale condensations from SMA 1.3 mm continuum emission of these MDCs. With a resolution of $\sim$0.2$\arcsec$($\sim$280 au at a distance of 1.4 kpc), the NOEMA 1.3 mm continuum data revealed varying levels of fragmentation within each source. For instance, some MDCs exhibit nine fragments around 300 au in radii (e.g., N30), while others show limited fragmentation (e.g., N56, NW14).

To identify fragments in the 1.3 mm continuum images, we employed the dendrogram algorithm \citep{2008ApJ...679.1338R}, implemented with astrodendro\footnote{\url{https://dendrograms.readthedocs.io/en/stable/}}. For 2D images, astrodendro requires three input parameters: $I_\mathrm{min}$, the minimum intensity threshold for pixels to be included in a structure; $\delta_\mathrm{min}$, the minimum difference in peak intensity required to distinguish two structures as separate structures; and $N_\mathrm{pix}$, the minimum number of pixels needed to define the smallest detectable structure. In our analysis, we set $I_\mathrm{min} = 4\sigma_\mathrm{rms}$, $\delta_\mathrm{min} = 2\sigma_\mathrm{rms}$, and $N_\mathrm{pix} \simeq 36$, corresponding to the size of the synthesized beam for each source. The leaf structures identified by astrodendro are treated as individual fragments. The effective radii of these fragments are calculated as $R_\mathrm{eff} = \sqrt{A/\pi}$, where $A$ is the exact area of the leaf structures derived from astrodendro. The radii of the extracted fragments range from 250 au to 820 au.

\begin{table*}[!ht]
    \centering
    \caption{Physical parameters of fragments.}
    \begin{tabular}{llllllllllllll}
    \hline
    \hline
    Fragment & Flux\tablefootmark{(a)} & $\mathrm{R_{eff}}$\tablefootmark{(b)} & $\theta_\mathrm{outflow}$\tablefootmark{(c)} & $\theta_\mathrm{gradient}$\tablefootmark{(d)} & $\upsilon_\mathrm{LSR}$\tablefootmark{(e)} & $T_\mathrm{gas}$\tablefootmark{(e)} & $M_\mathrm{gas}$ &  $M_*$\tablefootmark{(f)} & Q \\
     & (mJy) & (au) &  ($^\circ$) &  ($^\circ$) & ($\mathrm{km~s^{-1}}$) & (K) & ($M_\odot$) & ($M_\odot$) & \\
    \hline
    N30 e1 (VLA 3) & 172.5 & 455 &  120 & $55.4\pm7.0$ & $8.48\pm0.05$ & $254\pm13$ &  $0.44\pm0.023$ & $9.4\pm1.8$ & $5.2\pm0.6$ \\
    N30 e2 (VLA 1) & 177.6 & 821 & 28/65\tablefootmark{*} & $-157.7\pm1.0$ & $8.62\pm0.02$ & $114\pm3$ & $1.05\pm0.031$ &  \\
    N30 e5 (VLA 2) & 37.3 & 253 & 55 & ... & $11.54\pm0.03$ & $149\pm7$ & $0.17\pm0.009$ & \\
    N56 e2 & 40.1 & 343 & 50 & $-24.8\pm4.5$ & $21.35\pm0.17$ & $168\pm18$ & $0.17\pm0.019$ &  &  \\
    N63 e1 & 649.9 & 621 & 60\tablefootmark{*}/100/120/ & $132.8\pm2.4$ & $-5.20\pm0.05$ & $137\pm11$ & $3.17\pm0.298$  \\
    & & & 140\tablefootmark{*}/170\tablefootmark{*} & \\
    N68 e1 & 21.0 & 319 & 90 & ... & $-6.91\pm0.17$ & $173\pm39$  & $0.08\pm0.018$ \\
    NW14 e1 & 108.3 & 602 & 115 & $3.7\pm2.1$ & $5.26\pm0.05$ & $180\pm11$ & $0.40\pm0.025$ & $4.0\pm0.7$ & $3.9\pm0.4$ \\
    \hline
    
    \end{tabular}
    \tablefoot{
    \tablefootmark{a} 1.3 mm dust continuum flux density with the fragment after subtracting the contribution from free-free emission (see Section \ref{subsec:tgas_rhogas}). \tablefootmark{b} Effective radii of the fragments $R_\mathrm{eff}=\sqrt{A/\pi}$ (see Section \ref{subsec:dustcont}). \tablefootmark{c} Position angle for identified CO outflow axis. \tablefootmark{*} indicates the orientation of the unipolar CO outflow associated with the fragment. \tablefootmark{d} Position angle of velocity gradient (the direction of increasing velocity, measured east of north) detected in $\mathrm{CH_3CN~(12_3-11_3)}$ emission. ``...'' means no clear velocity gradients were detected. \tablefootmark{e} Fitting results of $\mathrm{CH_3CN}$ emission average over the region of identified fragments in Section \ref{subsec:tgas_rhogas}. \tablefootmark{f} The central enclosed mass by fitting the Keplerian profile to the PV diagram of $\mathrm{CH_3CN~(12_3-11_3)}$ using \emph{KeplerFit}. }
    \label{tab:frag_proper}
\end{table*}

\subsection{Dense gas kinematics}\label{subsec:kinematics}
Several transitions of dense gas tracers (e.g., $\mathrm{CH_3CN}$, $\mathrm{CH_3OH}$, etc.) were detected in these cores in the SMA observations in the 1.3 mm band~\citep{Pan2024}. $\mathrm{CH_3CN}$ has proven to be an excellent tracer of dense gas. This symmetric top molecule emits rotational spectra with a ladder of components (known as the \emph{K}-ladder) closely spaced in frequencies. The level populations of the K components can yield estimates of gas temperatures and densities (see Section \ref{subsec:tgas_rhogas}). Although the $\mathrm{CH_3CN}$ and $\mathrm{CH_3OH}$ emission trace dense gas typically seen in disks, the high-excitation transitions can also be influenced by molecular outflows \citep[e.g.,][]{2017ApJ...847...87S,2019ApJ...873...73Z}.

\begin{figure*}[!htp]
    \centering
    \includegraphics[width=0.95\textwidth]{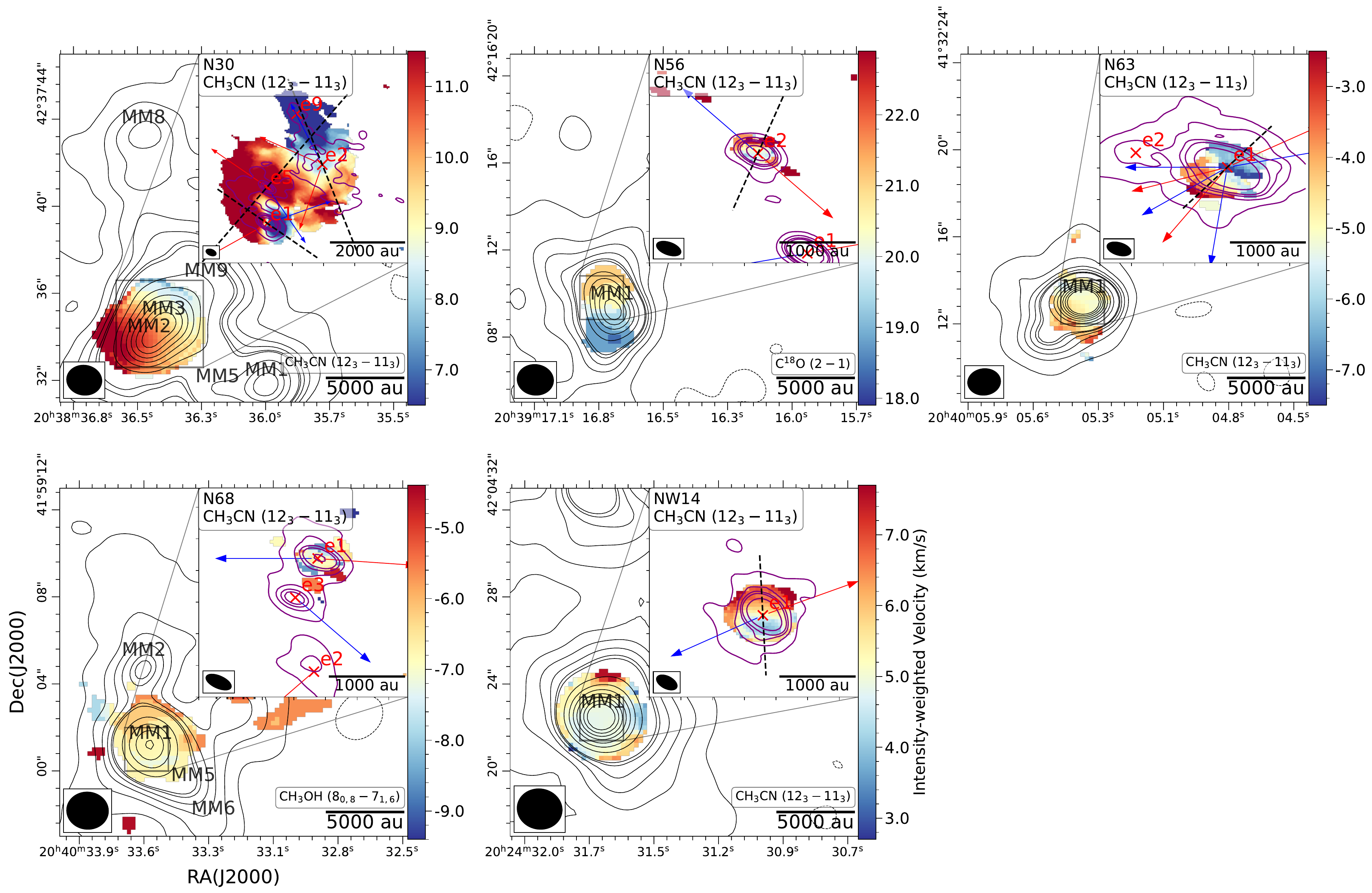}
    \caption{Zooming in on the velocity fields of each massive dense core. The main panel displays intensity-weighted mean velocity maps of the molecular line emission observed by SMA (see Fig. 3 in \cite{Pan2024}), with the SMA 1.3 mm continuum emission overlaid in contours. In the top-right corner of each panel, the velocity field of the $\mathrm{CH_3CN~(12_3-11_3)}$ line from the NOEMA observations is shown, overlaid with the NOEMA 1.3 mm continuum emission in contours, as in Fig. \ref{fig:1.3mm_cont}. The blue and red arrows starting from the center of the fragments indicate the blueshifted and redshifted CO outflow axes, staring from the center of the corresponding fragment, respectively. The dashed black lines mark the orientations of the PV cuts shown in Fig. \ref{fig:pv_map}. The ellipses at the bottom-left corner of each image shows the synthesized beam for the corresponding observation.}
    \label{fig:ch3cn_kinematics}
\end{figure*}

Figures \ref{fig:ch3cn_kinematics} and \ref{fig:ch3oh_kinematics} show the intensity-weighted mean velocity (first-moment) maps of the $\mathrm{CH_3CN~(12_3-11_3)}$ and $\mathrm{CH_3OH~(8_{0,8}-7_{1,6})}$ emission for the five MDCs. It is worth noting that N56 e2 and N68 e1 exhibit detections of several transitions of $\mathrm{CH_3CN}$, including $\mathrm{CH_3CN~(12_3-11_3)}$, whereas none of these transitions were detected in the previous SMA observations. The dense gas tracers reveal clear velocity gradients in some fragments. To obtain the direction of the velocity gradient, we fit the velocity structures to the function \citep[following][]{Goodman1993}:

\begin{equation}
\upsilon_\mathrm{LSR}=\upsilon_0+\upsilon_\alpha\Delta\alpha+\upsilon_\delta\Delta\delta,
\end{equation}
where $\upsilon_0$ is the systemic velocity, $\Delta\alpha$ and $\Delta\delta$ are offsets in right ascension and declination, and $\upsilon_\alpha$ and $\upsilon_\delta$ are the velocity gradients in the $\alpha$ and $\delta$ directions, respectively. We can derive the position angle (P.A.) of the velocity gradient by:

\begin{equation}
    \theta_\mathrm{gradient} = \tan^{-1}\frac{\upsilon_\delta}{\upsilon_\alpha}
\end{equation}

These velocity gradients in the fragments revealed by different tracers ($\mathrm{CH_3CN}$ and $\mathrm{CH_3OH}$) are consistent with each other. The position angles of the velocity gradients revealed by the $\mathrm{CH_3CN~(12_3-11_3)}$ emission are listed in Table \ref{tab:frag_proper}. The gradient in N63 e1 and NW14 e1 aligns closely ($\delta\mathrm{PA}<15^\circ$) with those observed in the previous SMA data \citep[see Fig. 3 and Table 3 in][]{Pan2024}{}, while the velocity gradient in N56 e2 traced by the $\mathrm{CH_3CN~(12_3-11_3)}$ and $\mathrm{CH_3OH~(8_{0,8}-7_{1,6})}$ emission at a resolution of 0.2$\arcsec$ is about 30$^\circ$ offset from the velocity gradient traced by the $\mathrm{C^{18}O~(2-1)}$ emission from the SMA data at a resolution of 1.8$\arcsec$. In addition, previous SMA observations of the $\mathrm{CH_3CN}$ emission in N30 revealed a clear velocity gradient along the southeast-northwest direction spanning 0.05 pc, while the higher resolution NOEMA data revealed a more complex velocity structure across the three identified fragments (N30 e1, e2, e5). These complex velocity patterns in N30 will be discussed in detail in Section \ref{subsec:disk}. For N68, the emission of the $\mathrm{CH_3CN~(12_3-11_3)}$ transition does not show a clear velocity gradient.

\subsection{Molecular outflows}
\begin{figure*}[!htp]
    \centering
    \includegraphics[width=0.95\textwidth]{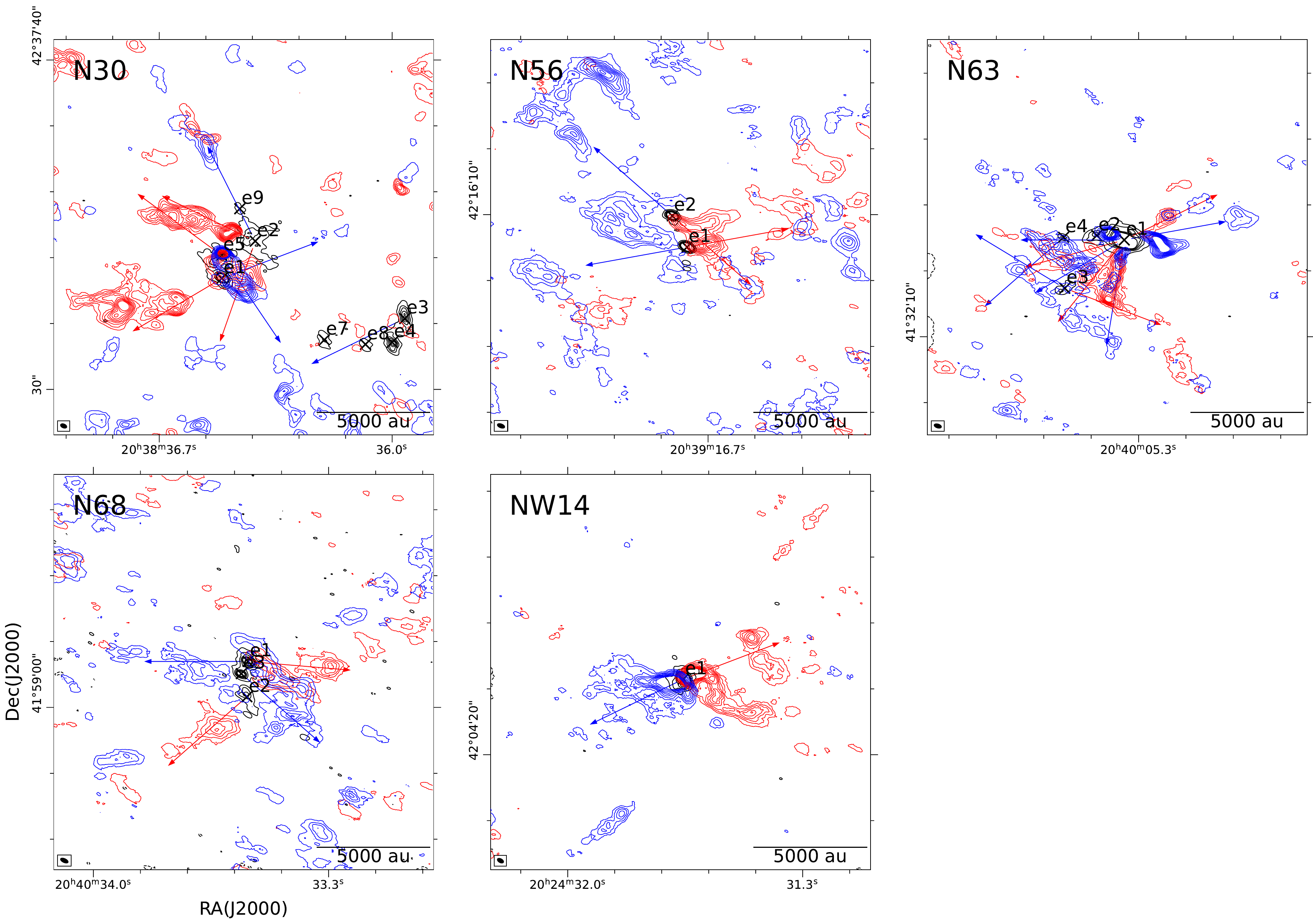}
    \caption{Integrated CO 2-1 emission in each source, overlaid with the NOEMA 1.3 mm continuum emission in black contours, the same as in Fig. \ref{fig:1.3mm_cont}. The red and blue contours show the redshifted and blueshifted CO emission in outflows, respectively. The contour levels start from 3$\sigma$ and increase in steps of 3$\sigma$, where $\sigma$ is the rms of the integrated CO emission. The red and blue arrows mark the identified outflow axes.}
    \label{fig:co_outflow}
\end{figure*}

A simple criterion to confirm whether the velocity gradient revealed by dense gas tracers (e.g., $\mathrm{CH_3CN}$, $\mathrm{CH_3OH}$) originates from a Keplerian disk is that the observed velocity gradient should be within 25$^\circ$ of being orthogonal to the outflow axis \citep{Cesaroni2007,Cesaroni2017}. In Fig. \ref{fig:co_outflow}, we present velocity-integrated intensity maps of the CO (2-1) emission. Since the CO emission close to the cloud velocity is typically spatially extended, and is largely filtered out by the interferometer, we integrate high-velocity CO (2-1) emission ($|\upsilon-\upsilon_\mathrm{sys}|\gtrsim3~\mathrm{km~s^{-1}}$, where $\upsilon_\mathrm{sys}$ is the systemic velocity of the central source) to minimize the filtering effects of an interferometer and to focus on the outflow gas. 

In Fig. \ref{fig:co_outflow}, most sources appear to have multiple outflows, potentially originating from different fragments within the dense core. In sparsely fragmented MDCs such as N56 and NW14, the CO outflow features are relatively clear, making it easier to identify the bipolar outflows associated with the 1.3 mm fragments. However, in highly fragmented sources such as N30 and N63, the outflow emission is more complex, making it challenging to determine the precise direction of molecular outflows. We therefore used our CO and SiO emission (shown in Fig. \ref{fig:sio_outflow}) in conjunction with other more detailed case studies for the sample to constrain outflow directions. 

As an example, N30, also known as W75N (B), contains a group of massive protostars \citep{2003ApJ...584..882S}, including VLA 1, VLA 2, VLA 3 \citep[e.g.,][]{1997ApJ...489..744T,2015Sci...348..114C}, which is associated with N30 e2, e5 and e1 in our sample, respectively. We followed the outflow directions identified by \cite{2003ApJ...584..882S}, which include a unipolar red-shifted CO outflow extending northeast of VLA 1 and a bipolar outflow extending to the southeast-northwest of VLA 3. They also found a unipolar outflow in high-velocity blue-shifted emission associated with MM1, which is further resolved into e3/e4 in our NOEMA data. The blue-shifted unipolar outflow is extending away from MM1 along the southeast direction, which is also detected in our NOEMA data with faint emission. \cite{2015Sci...348..114C} reported that VLA 2 is associated with a wind-driven $\mathrm{H_2O}$ maser shell, which has evolved from an almost isotropic outﬂow to a collimated one along the southwest-northeast direction in just 20 yr. Meanwhile, in our NOEMA data, we found a pair of red-shifted CO emissions extending to the northeast and blue-shifted CO emissions extending to the southwest from VLA 2, with projected extent of the outflow lobes ($L_\mathrm{lobe}$) of about 3000 au ($\sim$2.0\arcsec). To estimate the dynamical age of this pair of bipolar outflows, we divide the extent of the outflow with the maximum velocity ($\upsilon_\mathrm{max}=25~\mathrm{km~s^{-1}}$, estimated from the CO data) $t_d=L_\mathrm{lobe}/\upsilon_\mathrm{max}\approx6.6\times10^2~\mathrm{yr}$. This lower limit is consistent with the early phase of evolution of the protostar VLA 2. We suggest these emissions may originate from the bipolar outflow associated with VLA 2. Additionally, recent ALMA observations \citep{2023ApJ...956L..45G} revealed a compact SiO emission, which is consistent with a toroid and a wide-angle outflow along the northeast-southwest direction, supporting the presence of an outflow associated with this fragment. We use red and blue arrows to indicate redshifted and blueshifted outflow axes, respectively. To better illustrate the outflow direction, we also present channel maps of the CO emission in the Appendix \ref{app:outflow}. 

In summary, we identified a total of 28 outflow lobes and features, of which 20 comprise 10 identified bipolar outflows, while 8 outflow features have no identified bipolar counterpart.

\section{Discussions} \label{sec:discussion}
\subsection{Identification of Keplerian disks}\label{subsec:disk}
Recently, evidence of accretion disks around massive protostars has been accumulating \citep[see the review paper,][]{Beltran2016}{}{}. However, many large ($\gg2000$ au) disk candidates are not fully resolved and likely host smaller Keplerian disks ($\lesssim1000$ au) inside that become detectable with higher-resolution observations. Recent ALMA observations with sub-arcsecond resolutions have resolved these small ($\lesssim1000$ au) Keplerian disks around massive protostars \citep[e.g.,][]{Maud2018,Ilee2018,Ginsburg2018,Jimenez-Serra2020,Liu2020,Beltran2022}{}{}.

\cite{Pan2024} identified nine 2000-au-scale condensations in the Cygnus-X cloud hosting disk candidates using 1.3 mm observations with the SMA (spatial resolution $\sim2700$ au). Four of them (N30 MM2, N56 MM1, N63 MM1, NW14 MM1) are covered in our NOEMA observations. With the higher angular resolution ($\sim300$ au) NOEMA observations, these disk candidates are resolved into single or multiple fragments. By comparing the relative orientations of outflows and velocity gradients revealed by the dense gas tracer (e.g., $\mathrm{CH_3CN}$), we can determine whether the Keplerian disk really exist at smaller scales and identify which fragment hosts the Keplerian disk.

\begin{figure*}
    \centering
    \includegraphics[width=0.98\textwidth]{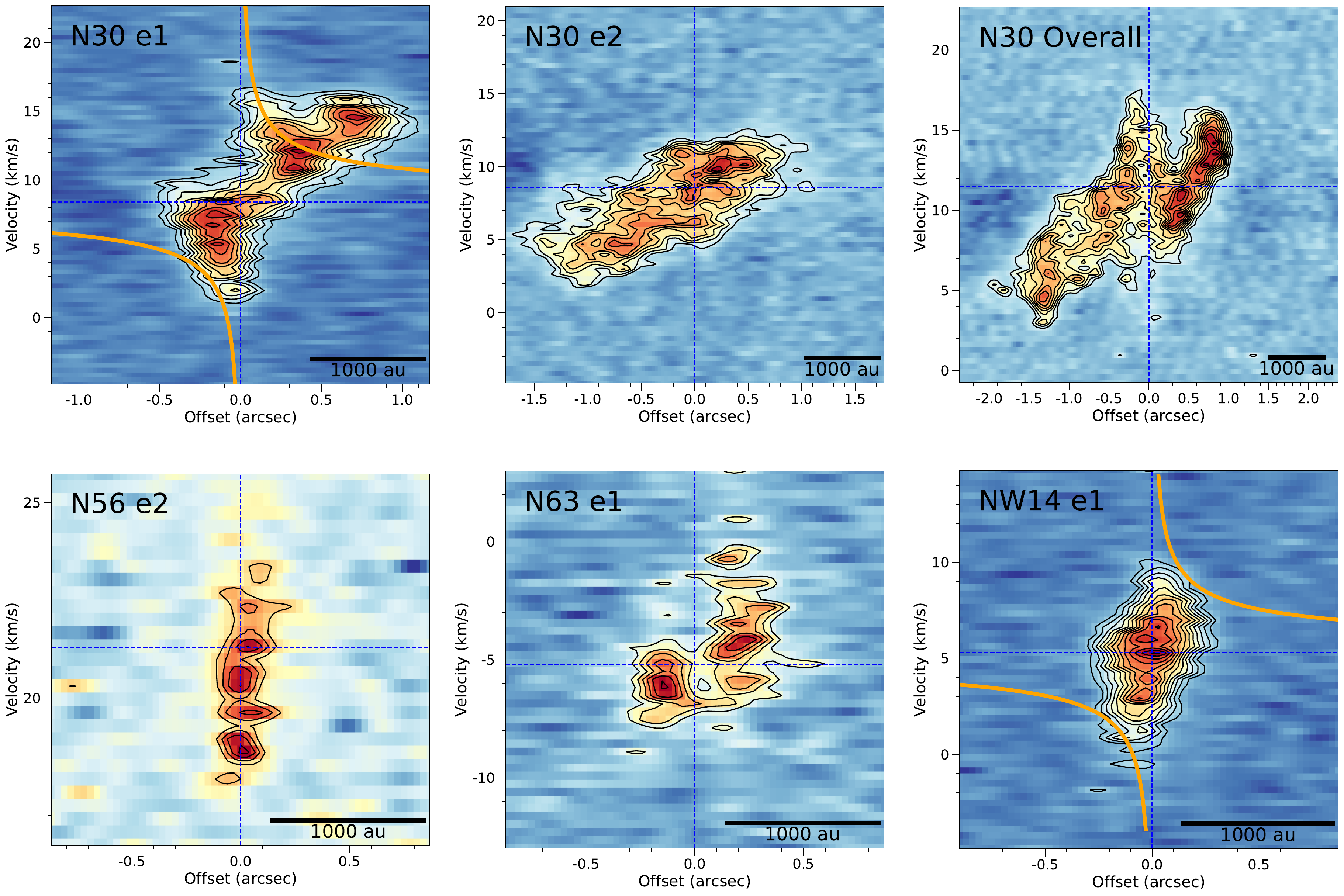}
    \caption{PV diagram of the $\mathrm{CH_3CN~(12_3-11_3)}$ line across fragments, cutting along the dashed black lines as shown in Fig. \ref{fig:ch3cn_kinematics}. The width of the cut is the size of a synthesized beam ($\sim0.18$\arcsec) and the reference position is the center of the corresponding fragment. Top left of each panel labels the name of the fragment. ``N30 overall'' indicates the PV plot centered at N30 e5 was cut along the northwest-southeast direction across the triple system in N30, as mentioned in Section \ref{subsec:disk}. The black contours start from $6\sigma$ and increase in steps of $3\sigma$. The orange solid lines show the Keplerian profile corresponding to the mass estimates listed in Table \ref{tab:frag_proper}. The vertical line marks the center of the corresponding fragment. The horizontal line indicates the systemic velocity derived from the $\mathrm{CH_3CN}$ line fitting of each fragment. The systematic velocity for ``N30 Overall'' is the fitting result of N30 e5.}
    \label{fig:pv_map}
\end{figure*}

We find that two sources in our sample, N56 e2 and NW14 e1, exhibit clear velocity gradients that are perpendicular to the outflow axis. This suggests that the velocity gradients traced by dense gas tracers likely originate from Keplerian disks. To better illustrate the gas kinematics, we present position-velocity (PV) diagrams along the direction of the velocity gradient (see Fig. \ref{fig:pv_map}). The PV diagram of NW14 e1 roughly resembles a Keplerian-like rotation curve ($\upsilon(r)\propto\sqrt{GM/r}$), while the excess emission in the quadrants opposite to the rotational motion may originate from the infalling material from the envelope \citep[e.g.,][]{Ohashi1997,Tobin2012} or outflowing gas. The disk radius of N56 e2 is approximately 343 au ($\sim0.2\arcsec$), which is only marginally resolved by our NOEMA data. Its PV diagram mimics a rigid-body-like rotation (i.e., $\upsilon(r)\propto r$), instead of a Keplerian rotation. However, synthetic observations by \cite{Ahmadi2019} show that PV diagrams of poorly resolved disks often resemble the rigid-body rotation. Therefore, observations with higher resolutions and sensitivity are needed to verify the presence of a Keplerian disk in N56 e2.

In N30, the dense gas emission mainly comes from the central region, which contains the massive protostars VLA 1, VLA 2, and VLA 3 (corresponding to e2, e5, e1, respectively). The velocity field of the $\mathrm{CH_3CN}$ and $\mathrm{CH_3OH}$ emission shows the greatest velocity differences along the northwest-southeast direction at scales of 3$\arcsec$ (about 4000 au) across the three fragments. This is consistent with the velocity field previously found in the SMA data using different dense gas tracers (e.g., $\mathrm{H_2CO}$, $\mathrm{CH_3CN}$) at scales of approximately 4$\arcsec$ and angular resolutions of 1$\arcsec$-2$\arcsec$ \citep[e.g., ][]{2021A&A...655A..86V,Zeng2023,Pan2024}. In particular, \cite{Pan2024} suggested that the velocity gradient arises from a 2000-au-scale disk surrounding one of the massive protostars, as it is perpendicular to the observed northeast-southwest CO outflow axis from the SMA data. 

With higher resolution NOEMA observations, we derive the PV diagram along the northwest-southeast (140$^\circ$) direction centered at N30 e5, as shown in the top-right panel in Fig. \ref{fig:pv_map}. It does not exhibit a butterfly pattern typically indicative of Keplerian rotation. Instead, it displays an "N-shape" pattern. The emission on the left side, ranging from 3 $\mathrm{km~s^{-1}}$ to 17 $\mathrm{km~s^{-1}}$, likely represents a gas flow extending from the parental envelope, channeling through N30 e2 to N30 e5. On the right side, the emission ranges from 16 $\mathrm{km~s^{-1}}$ to 8 $\mathrm{km~s^{-1}}$, suggesting a connection between the red-shifted emission from N30 e1 and N30 e5. Therefore, we think that the extended velocity gradient (about $4\arcsec$) probably traces gas flows connecting the three protostars, as already seen in other multiple systems through high-resolution observations and numerical simulations \citep[e.g.,][]{Krumholz2007,Tobin2016,Lee2023}{}{}. It is more clearly illustrated in the channel maps in Fig. \ref{fig:N30_ch3cn_chanmap}. 

In addition, we identify two distinct velocity gradients associated with e1 (VLA 3) and e2 (VLA 1). The velocity gradient associated with e1 is oriented in the northeast-southwest direction with a position angle of 55.4$^\circ$ nearly perpendicular to the axis of the outflow from e1 The gradient indicates the presence of a rotating structure, which is consistent with the $\mathrm{H_2CO}$ disk identified in \cite{2023ApJ...956L..45G} with comparable resolutions ($\sim0.15\arcsec$) using ALMA. The corresponding PV map in Fig. \ref{fig:pv_map} reveals a Keplerian-like disk pattern with additional red-shifted emission extending away from the disk, likely originating from the gas flow connecting e1 (VLA 3) and e5 (VLA 2). The estimate of enclosed mass of e1 is about 9.4 $M_\odot$ (see Section \ref{subsec:kepler_fitting}), which is consistent with the expected mass of an early-type B star for VLA 3 \citep{2001ApJ...546..345S,2003ApJ...584..882S}. It is also interesting to note that in the central region of N30 e1, we did not detect the $\mathrm{CH_3OH~(8_{0,8}-7_{1,6})}$ emission or any other $\mathrm{CH_3OH}$ transitions covered in our bandwidth. The explanation of the absence of the $\mathrm{CH_3OH}$ emission in N30 e1 is beyond the scope of this paper and will be discussed in a future work focusing on the chemistry of our NOEMA data. The velocity gradient associated with e2 is oriented the northeast-southwest direction with a position angle of -157.7$^\circ$. According to the channel map, the blue-shifted emission (from 2 $\mathrm{kms^{-1}}$ to 9 $\mathrm{kms^{-1}}$) extends to the northeast of e3 (VLA 1), stretching over an extent of 1.5$\arcsec$ (about 2100 AU). This structure is also reflected in the extended dust continuum emission. The PV map along this cut does not resemble the butterfly pattern characteristic of a Keplerian disk. It is possible that the velocity structures around e2 (VLA 1) are contaminated by the bipolar outflow, as the blue-shifted and red-shifted emission of $\mathrm{CH_3CN}$ around e2 (VLA 1) lies to the northeast and southwest, respectively, consistent with the orientation of the blueshifted and redshifted lobes of the CO outflows. 

Our 1.3 mm NOEMA observations have resolved N63 MM1 into several fragments. The emission from the dense gas tracers (e.g., $\mathrm{CH_3CN}$, $\mathrm{CH_3OH}$) mainly concentrates in e1, which shows a clear velocity gradient along the northwest-southeast direction. Previous SMA observations \citep[see Fig. 3. in][]{Pan2024} revealed a velocity gradient along the same direction in N63 MM1, which was identified as a disk candidate as the velocity gradient was roughly perpendicular to the bipolar outflow along the northeast-southwest axis. Our NOEMA observations have resolved the bipolar outflow into multiple uni/bipolar outflows. We identified ten outflow lobes and features, among which six comprise three identified bipolar outflows, while the other four outflow features do not have identifiable bipolar counterparts. Notably, the two identified bipolar outflows associated with e1 are both oriented approximately in the west-east direction, roughly aligned ( $\delta \mathrm{PA}\lesssim30^\circ$) with the velocity gradient traced by the $\mathrm{CH_3CN}$ and $\mathrm{CH_3OH}$ emission, suggesting that the velocity gradient may not originate from a Keplerian disk. Meanwhile, the PV plot (see Fig. \ref{fig:pv_map}) along the velocity gradient does not resemble the butterfly pattern characteristic of a Keplerian disk either. However, we cannot rule out the possibility of unresolved Keplerian disks within e1 due to the potential contamination of dense gas kinematics from multiple outflows in a potentially multiple system. In similar cases where other multiple uni/bipolar outflows have been found \citep[][]{2008ApJ...686L.107C,2015ApJ...798...61T,2019ApJ...870...81T,2022ApJ...927...54O}{}{}, subsequent observations with sufficient resolutions have revealed that the central protostellar source is a multiple system. Higher resolution ($\lesssim$100 au) observations are required to confirm the existence of disk candidates within N63 e1.

In N68 e1, we did not find a clear velocity gradient in the $\mathrm{CH_3CN~(12_3-11_3)}$ or $\mathrm{CH_3OH~(8_{0,8}-7_{1,6})}$ emission, which is consistent with the result from the low resolution SMA observations. However, since the signal-to-noise ratio (SNR) of the $\mathrm{CH_3CN~(12_3-11_3)}$ emission is limited (SNR$\sim$3, see Fig. \ref{fig:ch3cn_fitting_all}), observations with higher sensitivities may help to reduce the effect of noise on the gas kinematics traced by $\mathrm{CH_3CN}$. 

In summary, using higher resolution NOEMA data, we identified two small ($\sim$ 500 au) Keplerian-like disks in N30 e1, NW14 e1 out of the four large ($\gtrsim2000$ au) disk candidates identified from the SMA study.

\subsection{Dynamical mass estimates}\label{subsec:kepler_fitting}
To obtain the mass of central protostar, \cite{Seifried2016} proposed a method for determining the maximum rotation velocity ($\mathrm{\upsilon_{rot,max}}$) as a function of projected distance (\emph{x}) to the central protostar from PV diagrams and fitting a Keplerian profile to the $\upsilon_\mathrm{rot,max}(x)$ curve:

\begin{equation}
\upsilon(x)=\sqrt{G\frac{M_*+M_\mathrm{disk}}{x}}
\end{equation}

Synthetic ALMA observations of circumstellar disks in Keplerian rotation \citep[e.g.,][]{Seifried2016,Ahmadi2019}{}{} found that fitting the edge of emission in PV diagrams provides a more accurate estimate for the enclosed mass than other methods. However, it is also worth noting that this method only yield the lower limit to the central mass since it does not take the disk inclination into account. 

Using \emph{KeplerFit}\footnote{A python package to fit a Keplerian velocity distribution model to position-velocity data, in order to estimate mass of the central protostar.} package \citep{2019A&A...629A..10B}, we modeled the Keplerian profile at the outermost emission edges to estimate central protostar masses. We set the PV diagram center to the fragment centers identified by astrodendro, with central velocities based on $\mathrm{CH_3CN}~(12_K-11_K)K=0-6$ line fitting results  of $\mathrm{CH_3CN}~(12_K-11_K)~K=0-6$ lines (see Section \ref{subsec:tgas_rhogas}). Fits to the $4\sigma$ and $6\sigma$ emission edges of $\mathrm{CH_3CN(12_3-11_3)}$ yielded similar mass estimates, so we adopted the $6\sigma$ results. We generated PV plots for each fragment across multiple directions within the velocity gradient's uncertainty range. Despite varying orientations, the patterns and fitting results remain consistent, confirming the reliability of the derived velocity gradient and kinematic features and their insensitivity to directional uncertainties. Table \ref{tab:frag_proper} lists the best fitting results. The Keplerian profiles corresponding to the mass estimates are shown as the solid orange lines on the PV diagrams in Fig. \ref{fig:pv_map}.

\subsection{Gas temperatures and masses}\label{subsec:tgas_rhogas}
To estimate the physical properties of the dense gas traced by the $\mathrm{CH_3CN}$ emission, we averaged the $\mathrm{CH_3CN}$ spectra within astrodendro-identified fragments. We used the Spectral Line Identification and Modeling (SLIM) tool within the MADCUBA\footnote{MAdrid Data CUBe Analysis (MADCUBA) is a software developed in the Center of Astrobiology of Madrid (CSIC-INTA) to analyze astronomical datacubes and multiple spectra from the main astronomical facilities; \url{https://cab.inta-csic.es/madcuba/index.html}} package \citep{2019A&A...631A.159M} to fit the emission of $\mathrm{CH_3CN}~(12_K-11_K)~K=0-6$. We assumed a local thermodynamic equilibrium (LTE) analysis using SLIM, which is typically the case in such dense environments. To perform the LTE analysis, we left the column density ($N$), excitation temperature ($T_\mathrm{ex}$), the velocity ($\mathrm{\upsilon_{LSR}}$), the full width half maximum (FWHM), and the source size parameter (\emph{S}) as free parameters. We use the fitting results of gas velocity and temperature from \cite{Pan2024} as the initial guesses. For sources with a low signal-to-noise ratio (e.g., N63 e1), we found it necessary to fix one parameter to ensure the convergence of the fitting algorithm. The results of the LTE fits are summarized in Table \ref{tab:madcuba_fitting}.

\begin{figure}
    \centering
    \includegraphics[width=0.98\linewidth]{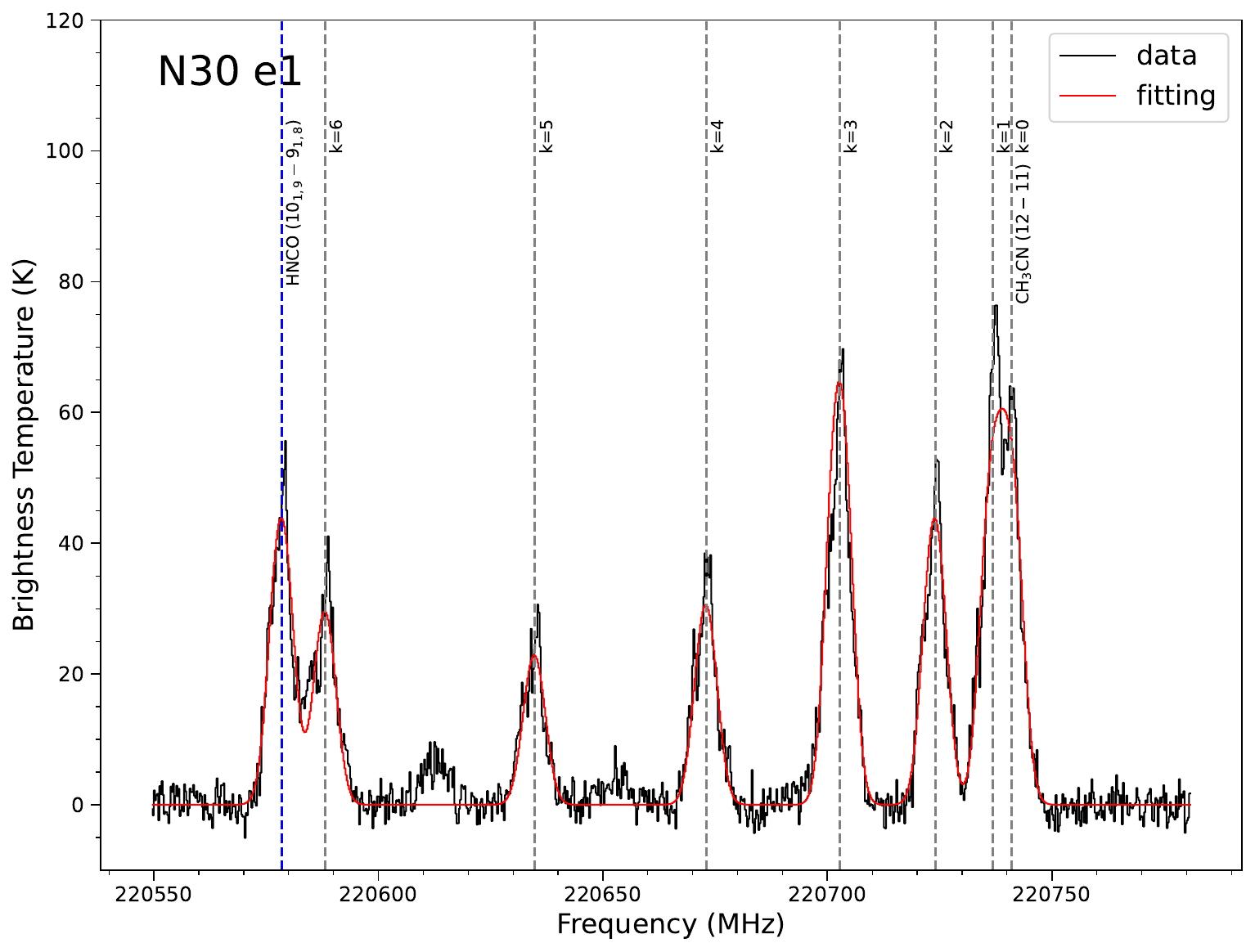}
    \caption{$\mathrm{CH_3CN}~(12_K-11_K)~K=0-6$ and $\mathrm{HNCO~(10_{1,9}-9_{1,8})}$  emission toward N30 e1. The red curves show the best LTE fit obtained with MADCUBA. Different transitions of $\mathrm{CH_3CN}$ are marked by grey dashed line. Blue dashed line represents the transition of $\mathrm{HNCO~(10_{1,9}-9_{1,8})}$.}
    \label{fig:ch3cn_fitting_example}
\end{figure}

We found that the $\mathrm{HNCO~(10_{1,9}-9_{1,8})}$ transition is also covered in the frequency range of $\mathrm{CH_3CN}~(12_K-11_K)~K=0-6$ and usually shows strong emission, which could affect the fitting results if we only consider the $\mathrm{CH_3CN}$ emission. To increase the fitting accuracy, we fitted HNCO and $\mathrm{CH_3CN}$ emission simultaneously, assuming the velocity and gas temperature of HNCO are the same as those of $\mathrm{CH_3CN}$. Fig. \ref{fig:ch3cn_fitting_example} and \ref{fig:ch3cn_fitting_all} shows the fitting results of all fragments detected in the $\mathrm{CH_3CN}~(12_K-11_K)~K=0-6$ emission. The physical properties derived by MADCUBA for each fragment are summarized in Table \ref{tab:frag_proper}. We find that the gas temperatures of these fragments consistently exceed 100 K. Numerical simulations \citep[e.g.,][]{Oliva2020,Oliva2023} demonstrate that fragmenting disks around massive protostars having mass of 10 $M_\odot$ exhibit temperatures of around 100 K at a radius of 500 au initially, which increase over time to 200 K. This temperature range aligns well with the values derived from the $\mathrm{CH_3CN}$ line fitting, suggesting that these fragments are conducive to the formation of high-mass protostars. Additionally, the column densities of $\mathrm{CH_3CN}$ ranges from $10^{14}$ to $10^{16}~\mathrm{cm^{-2}}$, which is consistent with those observed in other massive star forming regions \citep[e.g.,][]{Hernandez2014,Ahmadi2023}.

Assuming optically thin dust emission at 1.3 mm, we can derive the gas masses of these fragments through:

\begin{equation}
\centering
M_{\mathrm{gas}}=\frac{gF_\nu D^2}{B_\nu(T_d)\kappa_\nu},
\end{equation}

where $M_\mathrm{gas}$ is the gas mass of the condensation, g is the gas-to-dust mass ratio,  $F_\nu$ is the dust flux density at frequency, $\nu$, D is the source distance that is set to 1.4 kpc, $B_\nu(T_d)$ is the Planck function at a dust temperature, $T_\mathrm{d}$, and $\kappa_\nu$ is the dust opacity. We adopt the gas-to-dust mass ratio of 100 and the dust opacity $\kappa_\nu=0.899~\mathrm{cm^2~g^{-1}}$ \citep{Ossenkopf1994}. We assume a thermal equilibrium between gas and dust ($T_\mathrm{d} \approx T_\mathrm{gas}$) and use the gas temperature derived from $\mathrm{CH_3CN}$ fitting results to calculate the gas mass for each fragment. However, it is important to note that the $\mathrm{CH_3CN}$ emission is likely biased toward tracing warmer regions, whereas the dust emission can also probe cooler materials. Consequently, disk mass estimates derived from gas temperatures should be regarded as lower limits. According to \cite{Wang2022}, most fragments have negligible free-free contributions ($\lesssim2\%$) at 1.3 mm, except N30 e1 (VLA 3), which shows strong radio emission (13.5 mJy at 44 GHz). For this fragment, we scale the flux by $\nu^{-0.1}$ to estimate the expected free-free emission at 1.3 mm, and subtract this contribution from the millimeter continuum emission. The estimated contribution of the free-free emission is about 10\%. The derived gas masses of fragments range from $0.08M_\odot$ to $3M_\odot$, which is consistent with the masses of other disks around massive protostars \citep[e.g.,][]{Beltran2014,Maud2018,Busquet2019}. Table \ref{tab:frag_proper} lists the derived gas masses for each fragment. To estimate the optical depth, we calculate the average $\mathrm{H_2}$ column density for each fragment, which is approximately $10^{24}~\mathrm{cm^{-2}}$. Using $\kappa_\nu = 0.899~\mathrm{cm^2~g^{-1}}$, this corresponds to an optical depth of 0.04 at 230 GHz, confirming that the optically thin assumption remains valid.

\subsection{Stability of the identified disks}
\cite{Shu1990} made an analytical study of gravitational instabilities in gaseous disks and found that an accretion disk is stable only if its mass is
relatively small compared to the mass of the central protostar ($M_\mathrm{gas}<0.3~M_*$). For more massive disks, they are likely to be self-gravitating, which may induce substructures like spiral arms and fragments as predicted by numerical simulation \citep[e.g.,][]{Kratter2010,Meyer2017,Meyer2018,Kuiper2018,Oliva2020}. In our sample, the two identified disks have masses relatively small, meaning that both disks are gravitationally stable.

Estimating the Toomre Q parameter \citep{Toomre1964} is a widely used method to assess the gravitational stability of a rotating disk. However, this approach assumes a thin disk in Keplerian rotation, conditions that may not fully apply to the two candidate Keplerian disks identified in our study. As discussed in Section \ref{subsec:disk}, the position-velocity (PV) patterns of these candidates suggest additional kinematic contributions, such as gas flows connecting neighboring protostars or infall motions, rather than pure Keplerian rotation. Despite these deviations, we argue that the Toomre Q parameter remains a useful first-order diagnostic for evaluating the stability of these systems, providing some insights into their dynamical state. The Toomre Q parameter is defined as

\begin{equation}
    Q=\frac{c_s\Omega}{\pi G\Sigma}
\end{equation}

This parameter quantifies the balance between the stabilizing force of thermal pressure (represented by the sound speed, $c_s$) and shear pressure (represented by the angular velocity, $\Omega$) against the local gravity of the disk (represented by the angular velocity, $\Sigma$). Assuming the gas is adiabatic, sound speed can be estimated by $c_s=\sqrt{\gamma k_BT/\mu m_H}$, where $\gamma=7/5$ is the adiabatic index, $k_B$ is the Boltzmann constant, $\mu=2.8$ is the mean molecular weight, and $m_H$ is the mass of the hydrogen atom. The surface density of the disk can be calculated as $\Sigma=M_\mathrm{gas}/\pi R^2$, where R is the effective radius of the disk, listed in Table \ref{tab:frag_proper}. Based on the position-velocity diagram, we find that the two disks are potentially undergoing Keplerian rotation. Thus, we can derive the angular velocity of the system as $\Omega=\sqrt{G(M_*+M_\mathrm{gas})/R^3}$, where $M_*$ is the dynamical mass of the central protostar estimated from the PV diagram of $\mathrm{CH_3CN}$ emission. Since we do not account for inclination angles, these dynamical mass estimates represent a lower limit to the central protostellar masses, and as a result, the estimated Q values also serve as a lower limit. Table \ref{tab:frag_proper} lists the Q values for the two disks. A value of $Q<1$ indicates that the disk is prone to fragmentation, whereas $Q>1$ suggests that the system is stable against gravitational collapse. Both disks have $Q>2$, indicating stability, which is consistent with our findings based on the mass ratio between the disk and the central protostar. However, it is worth noting that the Toomre analysis depends on the spatial resolution. Numerical simulations \citep[e.g.,][]{Ahmadi2019, Oliva2020}{} have shown that locally unstable substructures can still exist within a globally stable disk. Given that the effective radius of both disks are around 500 au, one still needs higher spatial resolution observations ($\lesssim$ 100 au) to confirm whether these disks are truly gravitationally stable without further fragmentation.

\subsection{Comparison with the former disk candidates in the SMA observations}

Using the SMA observations at a 1.8$\arcsec$ resolution, \cite{Pan2024} only found nine 2000-au-scale condensations with evidence of rotation out of 27 sources associated with outflows and detected in high-density tracers. Due to the coarse resolution, those disk candidates are only marginally resolved.

Here, using high-resolution ($\sim$0.2$\arcsec$) NOEMA observations, we confirmed the presence of two candidate Keplerian disks at scales of 500 au (N30 e1, NW14 e1) out of four previously identified disk candidates (N30 MM2, N56 MM1, N63 MM1, NW14 MM1) from \cite{Pan2024}. NW14 MM1 is resolved into a single fragment, which shares a similar velocity gradient orientation. In contrast, N30 MM2 is resolved into four fragments, two of them (e1, e2) exhibiting distinct velocity gradients. Based on PV diagrams along these gradients, only the northeast-southwest velocity gradient associated with N30 e1 can be attributed to Keplerian rotation. The southeast-northwest velocity gradient, which has a similar orientation and extension with that observed in N30 MM2 using the SMA observations, was previously thought to originate from a 2000-au-scale disk. However, the PV diagram and channel map of $\mathrm{CH_3CN~(12_3-11_3)}$ from NOEMA data suggests that it more likely originates from gas flow connecting the fragments within N30 MM2.

Regarding the other two disk candidates (N56 MM1, N63 MM1), although the velocity gradient of N56 e2 is approximately perpendicular to the outflow axis, the PV diagram of the marginally resolved disk in N56 e2 does not show evidence of Keplerian rotation at scales of 500 au. Higher resolution observations are required to confirm the presence of a disk in this source. For N63 MM1, we found that the fragment with the strongest emission, N63 e1, shares a similar velocity gradient orientation with its parental structure. The 0.1-pc-scale collimated outflow breaks down into multiple uni- and bipolar outflows associated with N63 e1, none of which are perpendicular to the velocity gradient. These outflows may originate from different sources unresolved by our NOEMA data, suggesting a densely clustered environment. Higher-resolution observations ($\lesssim$ 100 AU) are therefore needed to pinpoint the origin of each outflow and to determine if smaller-scale disks are present within N63 e1.
 
\section{Summary and conclusions}\label{sec:summary}

In this work, we present gas kinematics of five massive dense cores (MDCs) in the Cygnus-X molecular cloud complex, observed with the NOEMA interferometer at a resolution of 0.2$\arcsec$ (approximately 300 au at a distance of 1.4 kpc). This sample includes four 0.01-pc-scale disk candidates (N30 MM2, N56 MM1, N63 MM1, NW14 MM1) previously identified by the SMA observations at a resolution of 1.8$\arcsec$ (approximately 2700 au at 1.4 kpc) and one source (N68 MM1) without evidence of gas rotation at these scales. With the higher resolution continuum and line emission data, we are able to examine the existence of Keplerian disks at smaller scales and identify which fragments
confirm candidate disks.

Using our 1.3 mm NOEMA data, we resolved all the disk candidates from the SMA observations into multiple fragments, except for NW14 MM1. These fragments range in size from 250 to 800 au, confirming that the previously identified 2000-au-scale disk candidates are not fully resolved.

All the five MDCs show detection of the \emph{K-}ladder rotational components of the dense gas tracer $\mathrm{CH_3CN}~(12_K-11_K)~K=0-6$. The velocity gradient revealed by $\mathrm{CH_3CN~(12_3-11_3)}$ across two fragments (N30 e1, NW14 e1) are perpendicular to the direction of the bipolar CO outflows, and their PV diagrams resemble differential rotation, confirming that these fragments might host candidate Keplerian
disks at scales of 500 au. In contrast, the fragments in the remaining three MDCs do not show clear evidence of Keplerian-like disks at these scales.

By modeling the level populations of the $\mathrm{CH_3CN}~(12_K-11_K)~K=0-6$ lines under LTE conditions using MADCUBA, we estimate gas temperatures in the fragments ranging from 110 to 250 K. Assuming gas and dust temperatures are in equilibrium, we calculate gas masses of the fragments, which range from 0.08 $M_\odot$ to 3.2 $M_\odot$, after excluding the contribution of the free-free emission from the 1.3 mm continuum emission.

Fitting the 6$\sigma$ edges of the PV diagrams to the Keplerian curve, we estimate the dynamical masses of the central objects for the two candidate Keplerian disks. We further calculate the Toomre Q parameters for both candidate disks and find that they are gravitationally stable overall.

In conclusion, we resolve substructures inside previously reported 2000 au disk candidates by SMA observations, and confirm two candidate Keplerian disks with radii of about 500 au. Higher-resolution ($\lesssim500$ au) and higher-sensitivity observations are required to resolve the 2000-au-scale disk candidates and check the existence of smaller Keplerian disks.

\begin{acknowledgement}
    K.Q. acknowledges supports from National Natural Science Foundation of China (NSFC) grants 12425304 and U1731237, and National Key R\&D Program of China No. 2023YFA1608204 and No. 2022YFA1603100. X. P. acknowledges support from the Smithsonian Astrophysical Observatory (SAO) Predoctoral Fellowship Program. This work is based on observations carried out under project number W22BE with the IRAM NOEMA interferometer. IRAM is supported by INSU/CNRS (France), MPG (Germany), and IGN (Spain). 
\end{acknowledgement}

\bibliography{CENSUS_NOEMA_disk}{}
\bibliographystyle{aa} %

\clearpage

%




\begin{appendix}
\onecolumn
\section{Velocity field maps}
\begin{figure}[!htp]
    \centering
    \includegraphics[width=0.85\textwidth]{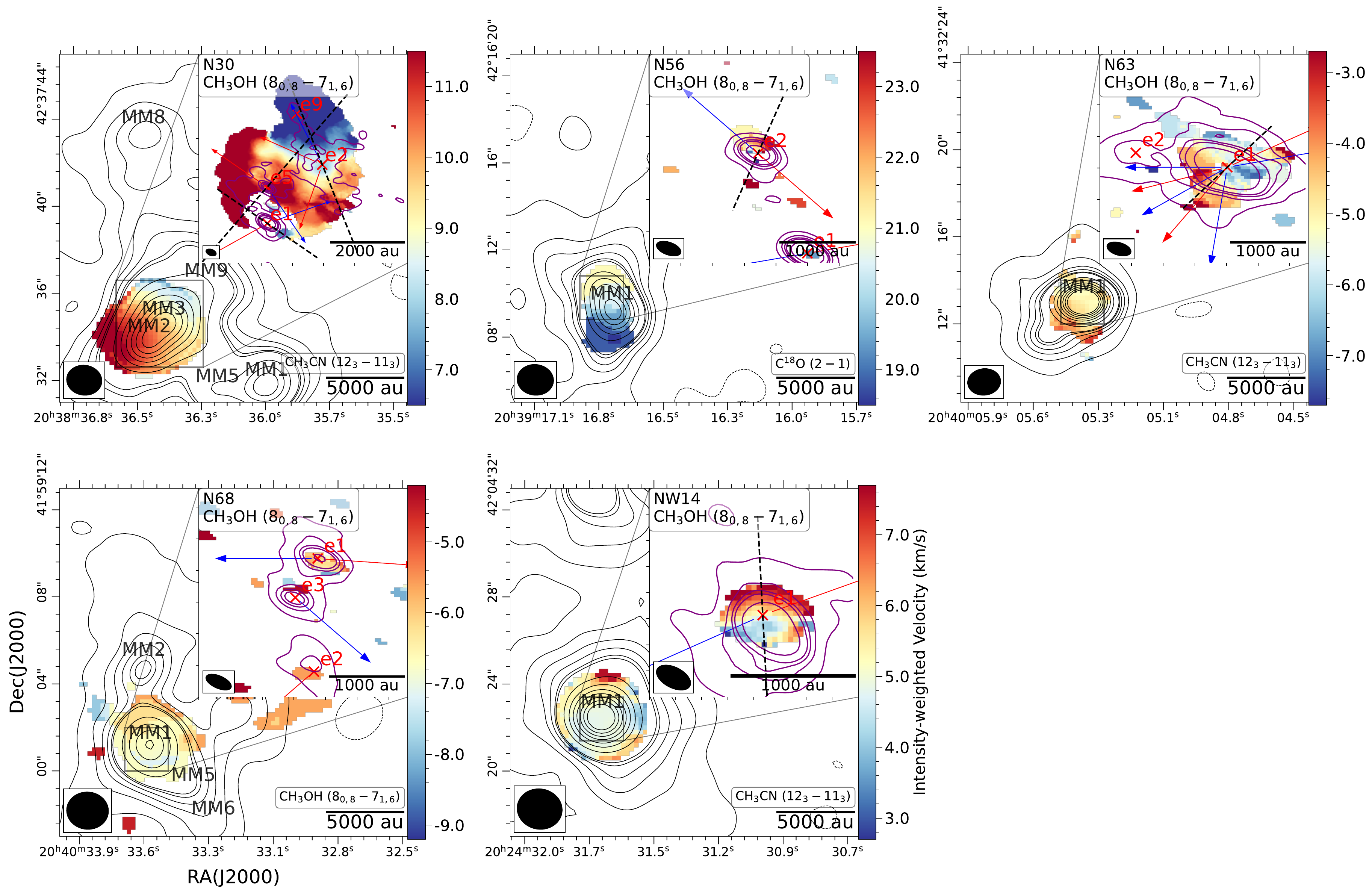}
    \caption{Same as Fig. \ref{fig:ch3cn_kinematics}, but for the $\mathrm{CH_3OH~(8_{0,8}-7_{1,6})}$ emission.}
    \label{fig:ch3oh_kinematics}
\end{figure}

\begin{figure}[!ht]
    \centering
    \includegraphics[width=0.8\textwidth]{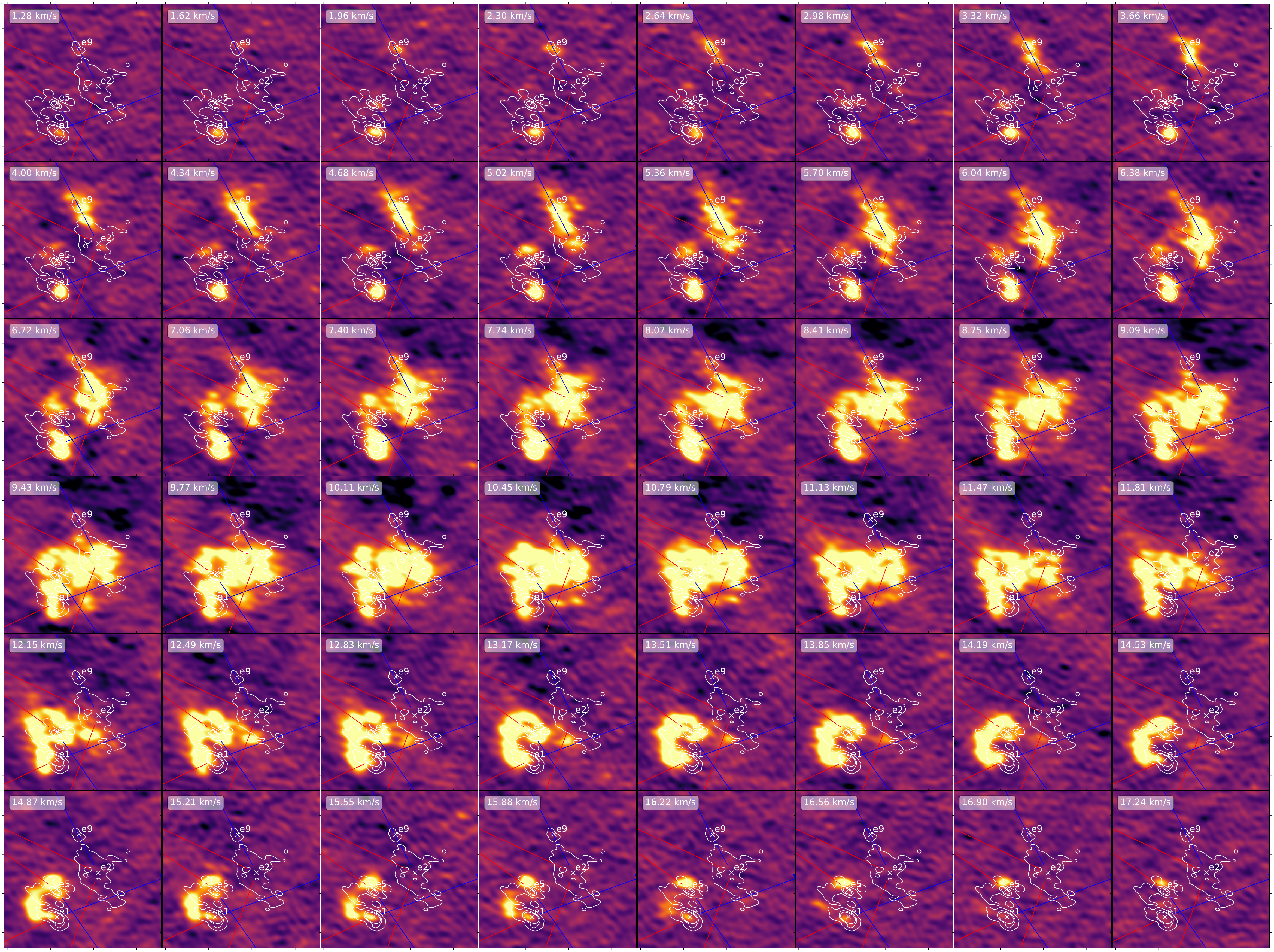}
    \caption{Channel maps for the $\mathrm{CH_3CN~(12_3-11_3)}$ emission in N30. The white contours represent the NOEMA 1.3 mm continuum emission same as in Fig. \ref{fig:1.3mm_cont}. Red and blue arrows represent the direction of red- and blue-shifted outflows, respectively.}
    \label{fig:N30_ch3cn_chanmap}
\end{figure}

\clearpage


\section{Molecular outflows}\label{app:outflow}
\begin{figure}[!ht]
    \centering
    \includegraphics[width=0.75\textwidth]{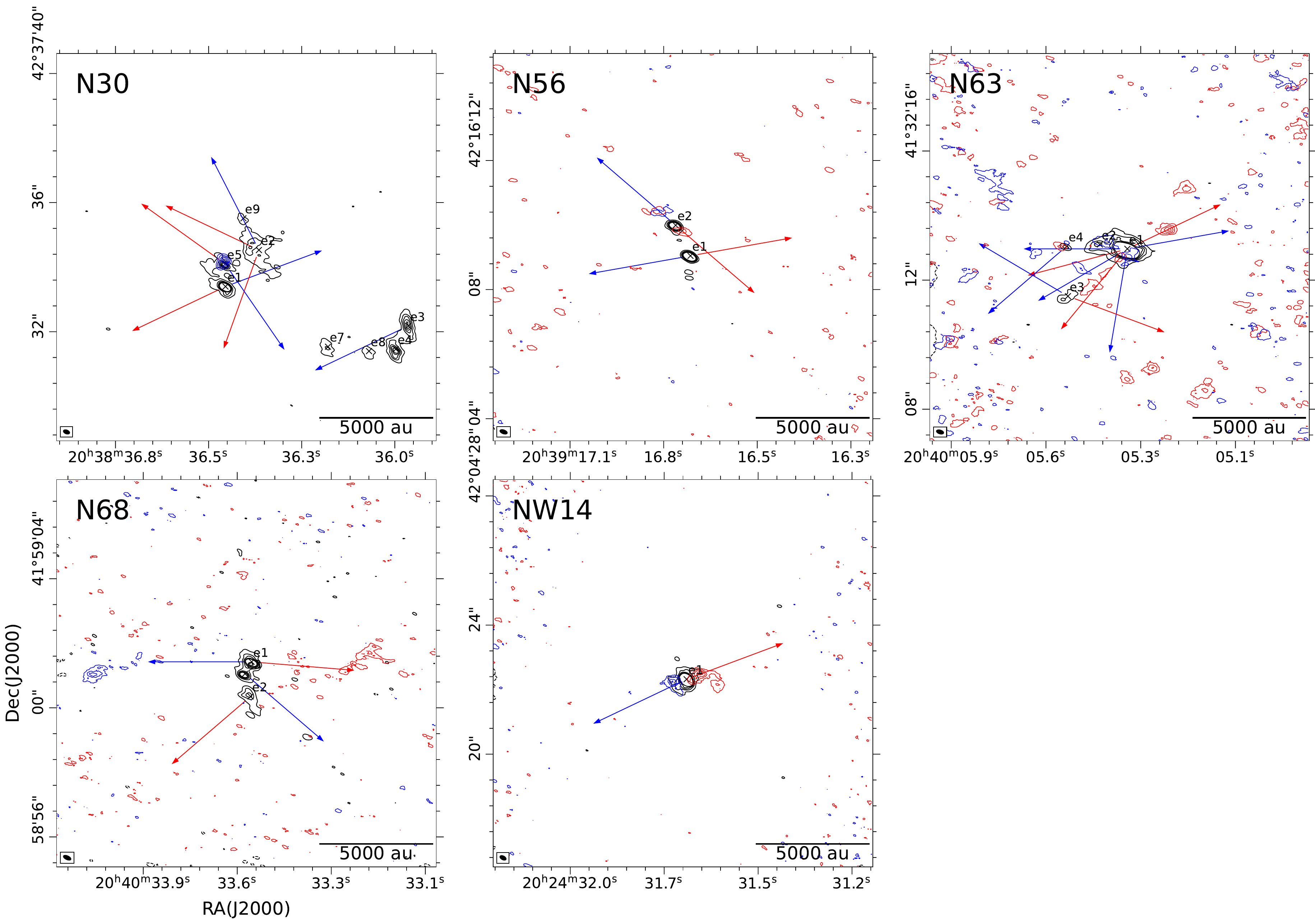}
    \caption{Velocity integrated emission of the SiO (5-4) transition.}
    \label{fig:sio_outflow}
\end{figure}

\begin{figure}[!ht]
    \centering
    \includegraphics[width=0.8\linewidth]{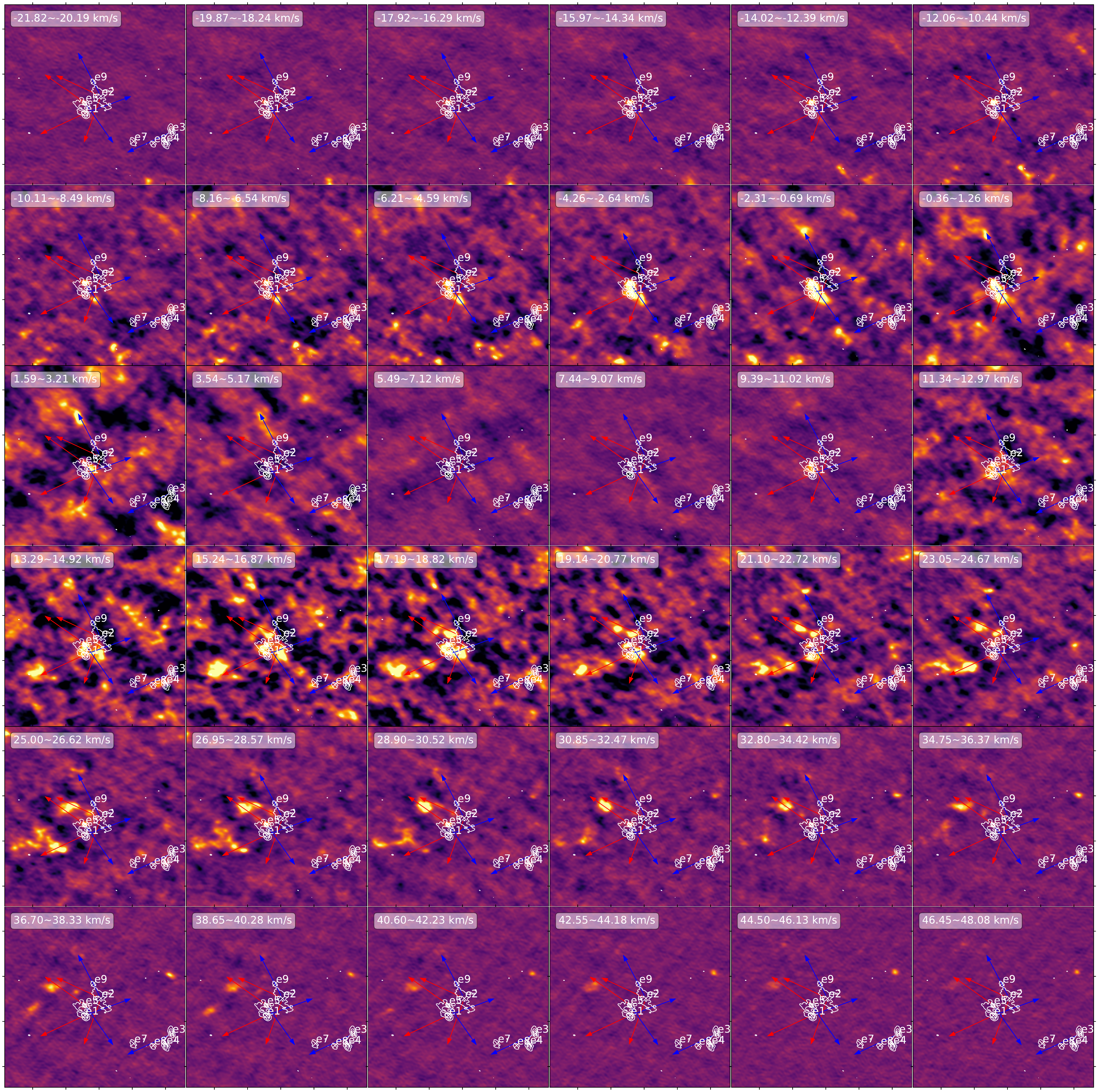}
    \caption{Channel maps for the CO (2-1) emission in N30. Each panel shows velocity integrated emission in a range of 1.5 $\mathrm{km~s^{-1}}$. The white contours represent the NOEMA 1.3 mm continuum emission same as in Fig. \ref{fig:1.3mm_cont}. Red and blue arrows represent the direction of red- and blue-shifted outflows, respectively.}
    \label{fig:N30_CO_chanmap}
\end{figure}

\begin{figure}[!ht]
    \centering
    \includegraphics[width=0.8\linewidth]{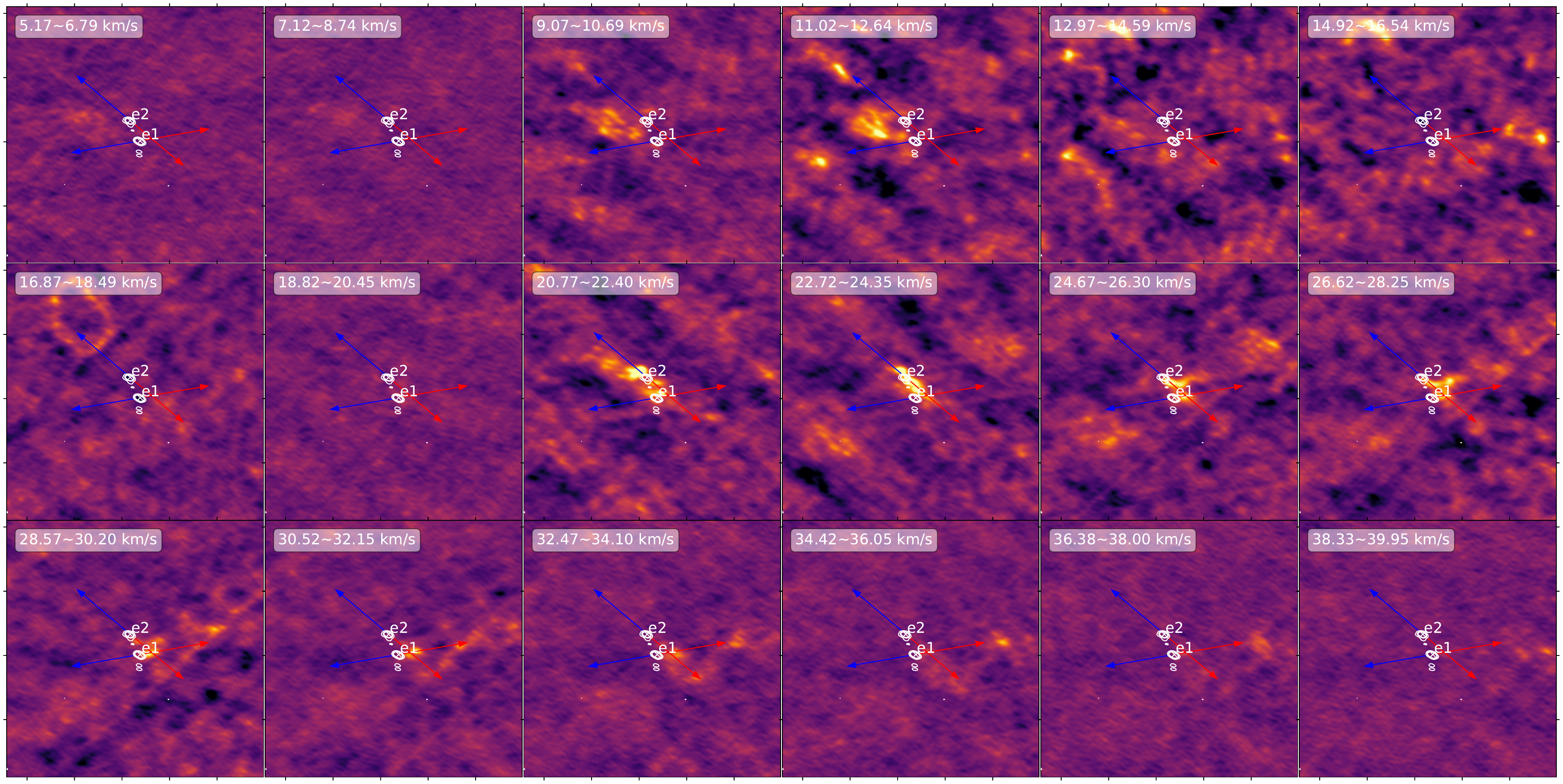}
    \caption{Same as Fig. \ref{fig:N30_CO_chanmap}, but for N56.}
\end{figure}

\begin{figure}[!ht]
    \centering
    \includegraphics[width=0.8\linewidth]{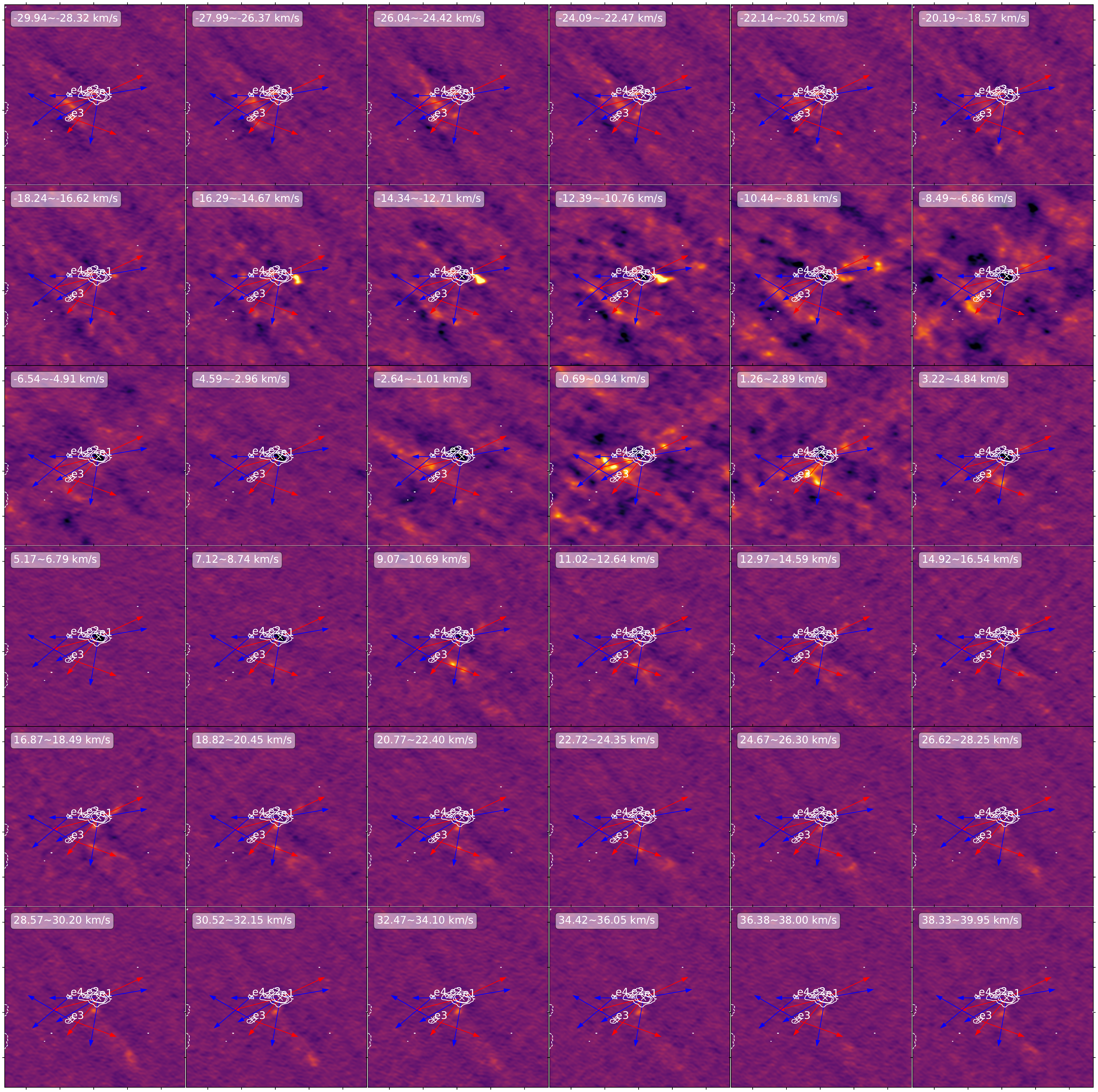}
    \caption{Same as Fig. \ref{fig:N30_CO_chanmap}, but for N63.}
\end{figure}

\begin{figure}[!ht]
    \centering
    \includegraphics[width=0.8\linewidth]{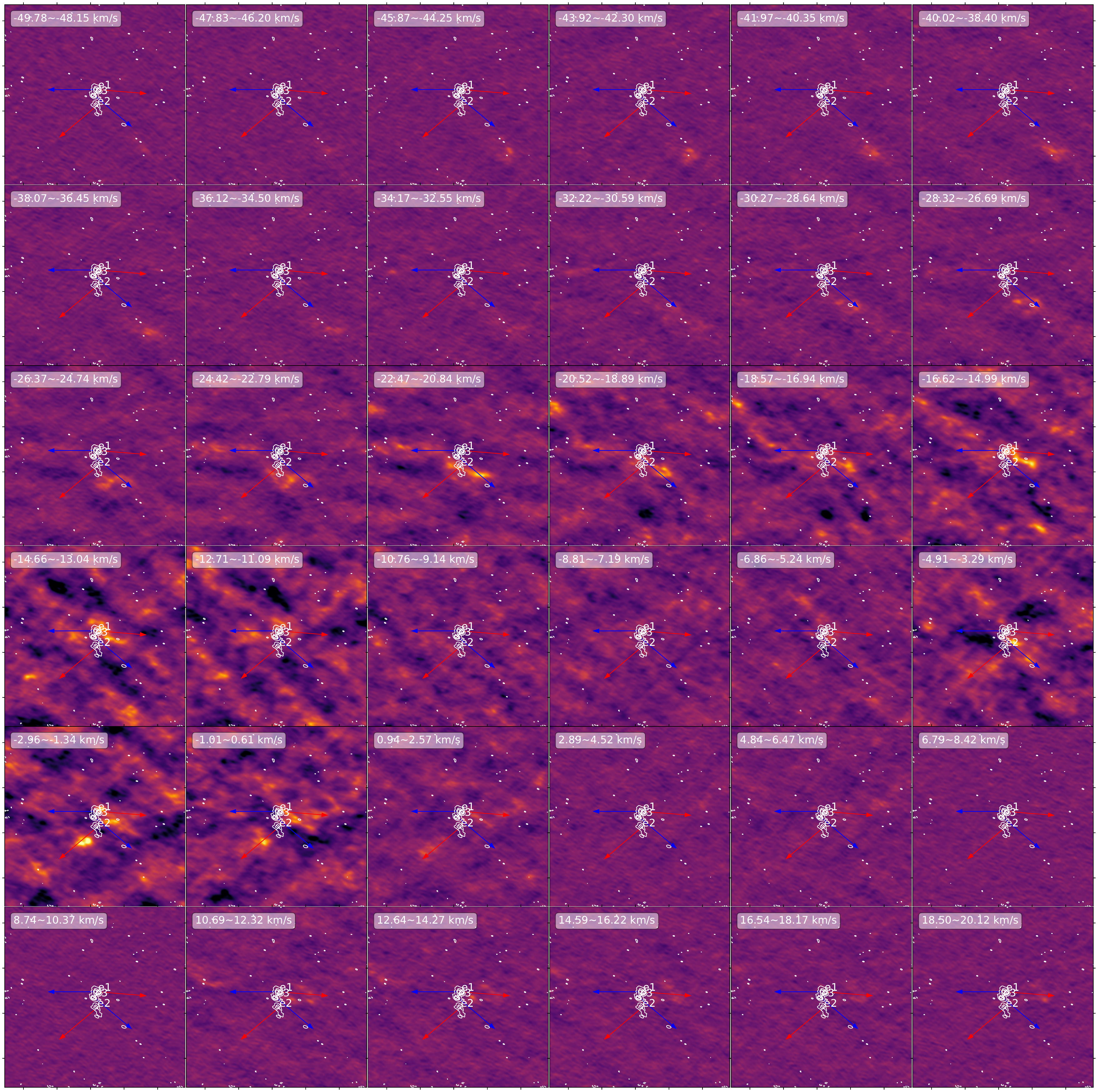}
    \caption{Same as Fig. \ref{fig:N30_CO_chanmap}, but for N68.}
\end{figure}

\begin{figure}[!ht]
    \centering
    \includegraphics[width=0.8\linewidth]{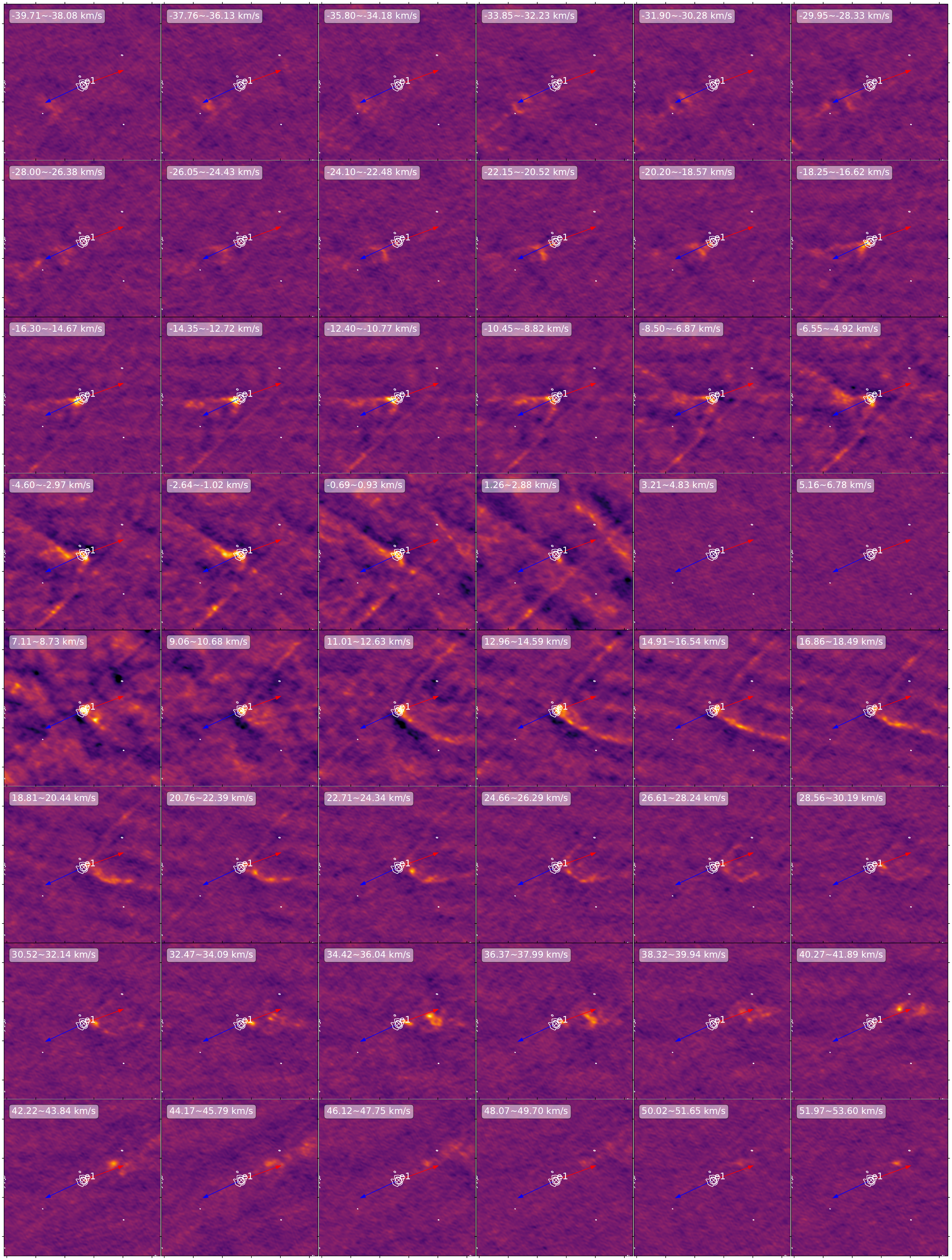}
    \caption{Same as Fig. \ref{fig:N30_CO_chanmap}, but for NW14.}
\end{figure}

\clearpage

\section{Line fitting result of $\mathrm{CH_3CN}$}
\begin{table}[!ht]
    \centering
    \caption{Fitting results of $\mathrm{CH_3CN}$}
    \begin{tabular}{lcccccc}
    \hline
    \hline
    Fragment & $T_\mathrm{gas}$ & $\upsilon_\mathrm{LSR}$ & $N$\tablefootmark{a} & FWHM\tablefootmark{b} & $S$ \tablefootmark{c} \\
    & (K) & ($\mathrm{km~s^{-1}}$) &($\times10^{15}~\mathrm{cm^{-2}}$) & ($\mathrm{km~s^{-1}}$) & (\arcsec) \\
    \hline
    N30 e1 & $254\pm13$ & $8.48\pm0.05$ & $27.2\pm3.6$ & $7.64\pm0.14$ & 0.0 \\
    N30 e2\tablefootmark{$\dagger$} & $114\pm3$ & $8.62\pm0.02$ & $23.0\pm0.6$ & $3.69\pm0.04$ & 0.20\\
    N30 e5\tablefootmark{$\dagger$} & $149\pm7$ & $11.54\pm0.03$ & $32.4\pm2.1$& 4.88\tablefootmark{*} & 0.16 \\
    N56 e2 & $168\pm18$ & $21.35\pm0.11$ & $1.0\pm0.2$ & $5.31\pm0.24$ & 0.0 \\
    N63 e1\tablefootmark{$\dagger$} & $137\pm11$ & $-5.20\pm0.05$ & $24.9\pm2.9$& 3.08\tablefootmark{*} & 0.08 \\
    N68 e1 & $173\pm39$ & $-6.91\pm0.17$ & $0.4\pm0.1$ & $4.04\pm0.35$ & 0.0 \\
    NW14 e1 & $180\pm11$ & $5.26\pm0.05$ & $16.4\pm2.0$ & $5.02\pm0.13$ & 0.0 \\
    \hline
    \end{tabular}
    \tablefoot{\tablefootmark{a} The column density of $\mathrm{CH_3CN}$, which is degenerated with source size. \tablefootmark{b} The full width half maximum derived from line fitting. \tablefootmark{*} Parameter fixed. \tablefootmark{c} The source size is derived from the line fitting analysis. A value of 0.0 indicates that the beam filling factor is equal to 1, meaning the source fully fills the beam. \tablefootmark{$\dagger$} means the components K=0-4 of $\mathrm{CH_3CN}$ are optically thick.}
    \label{tab:madcuba_fitting}
\end{table}
\begin{figure*}[!ht]
    \includegraphics[width=0.45\textwidth]{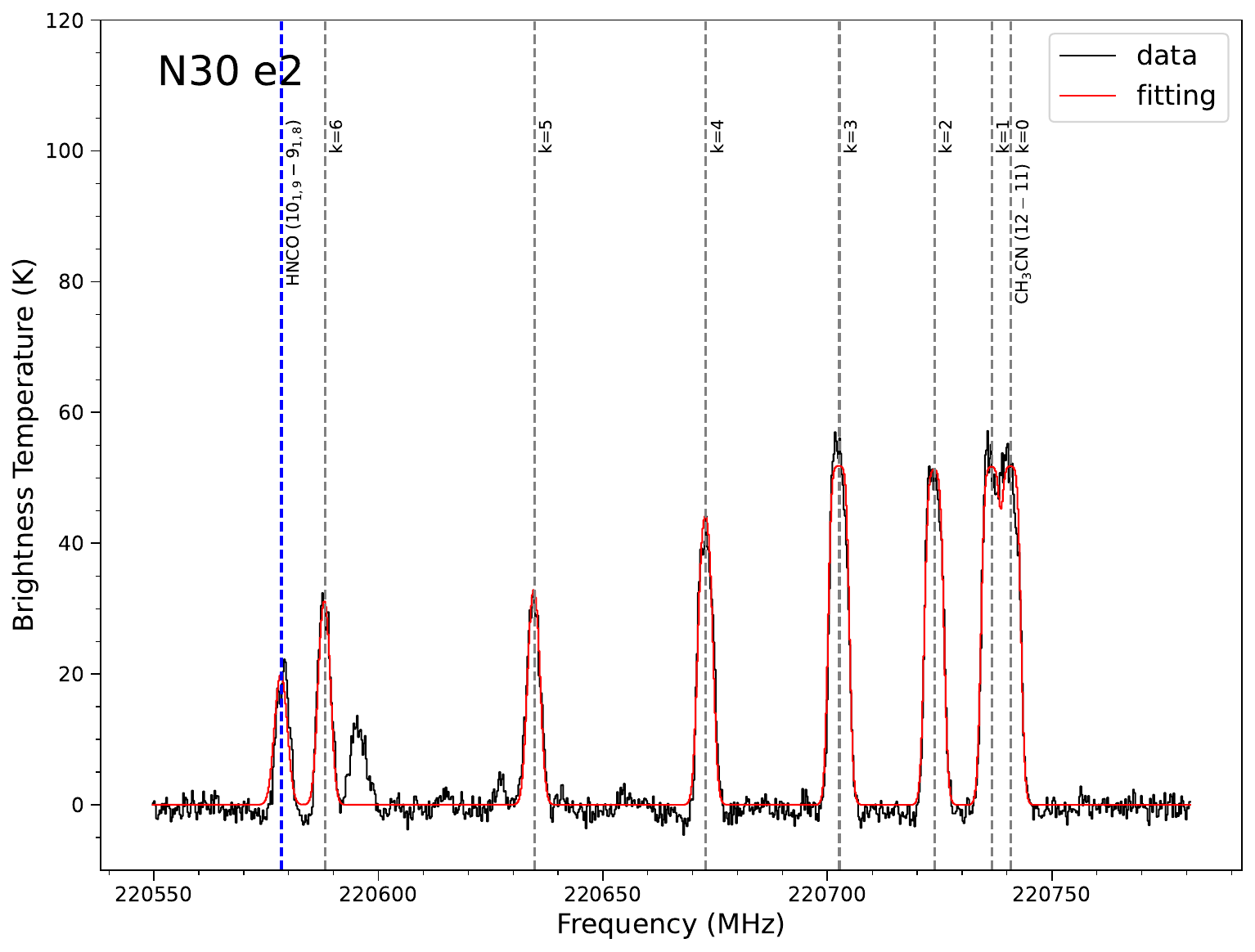}
    \includegraphics[width=0.45\textwidth]{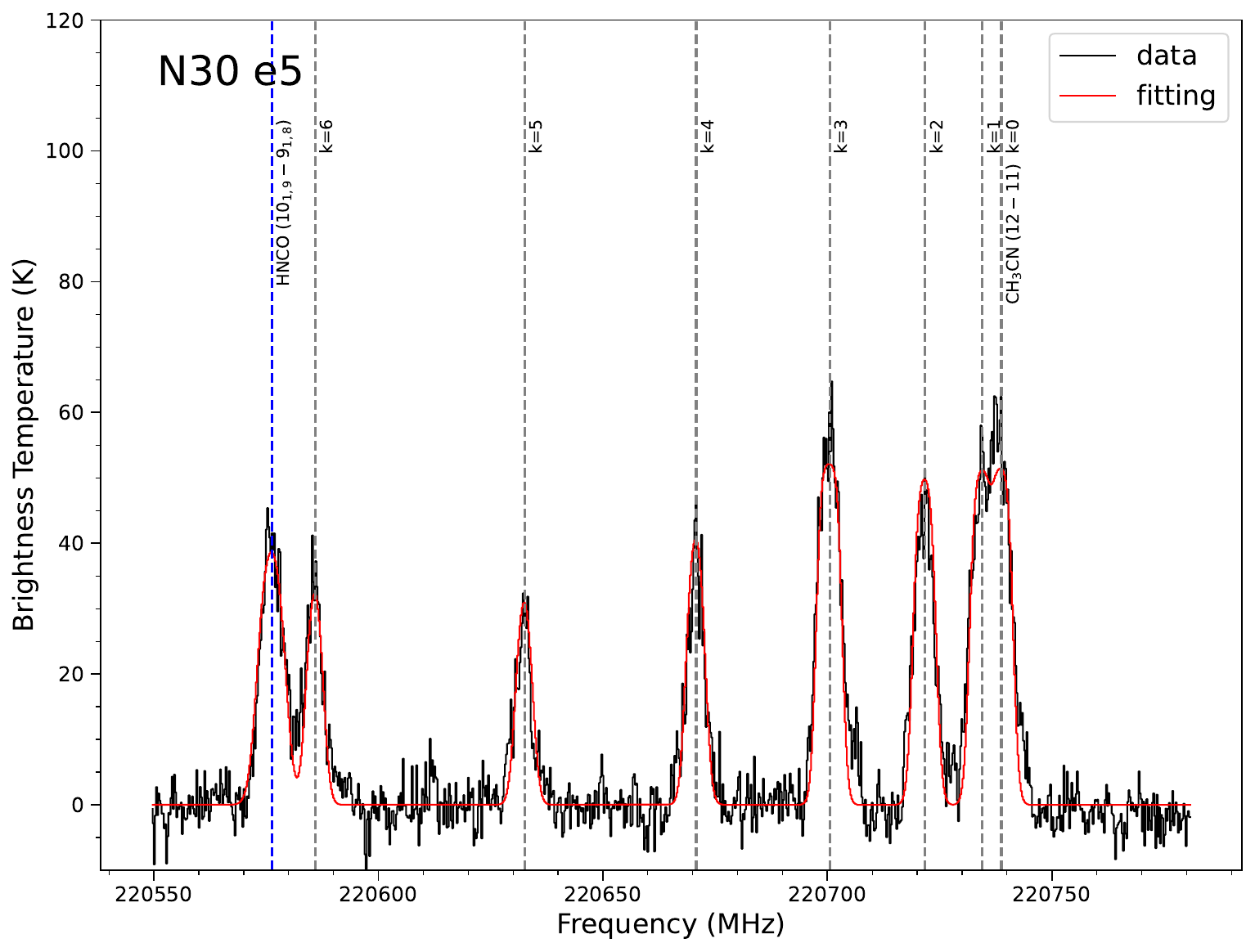}\\
    \includegraphics[width=0.45\textwidth]{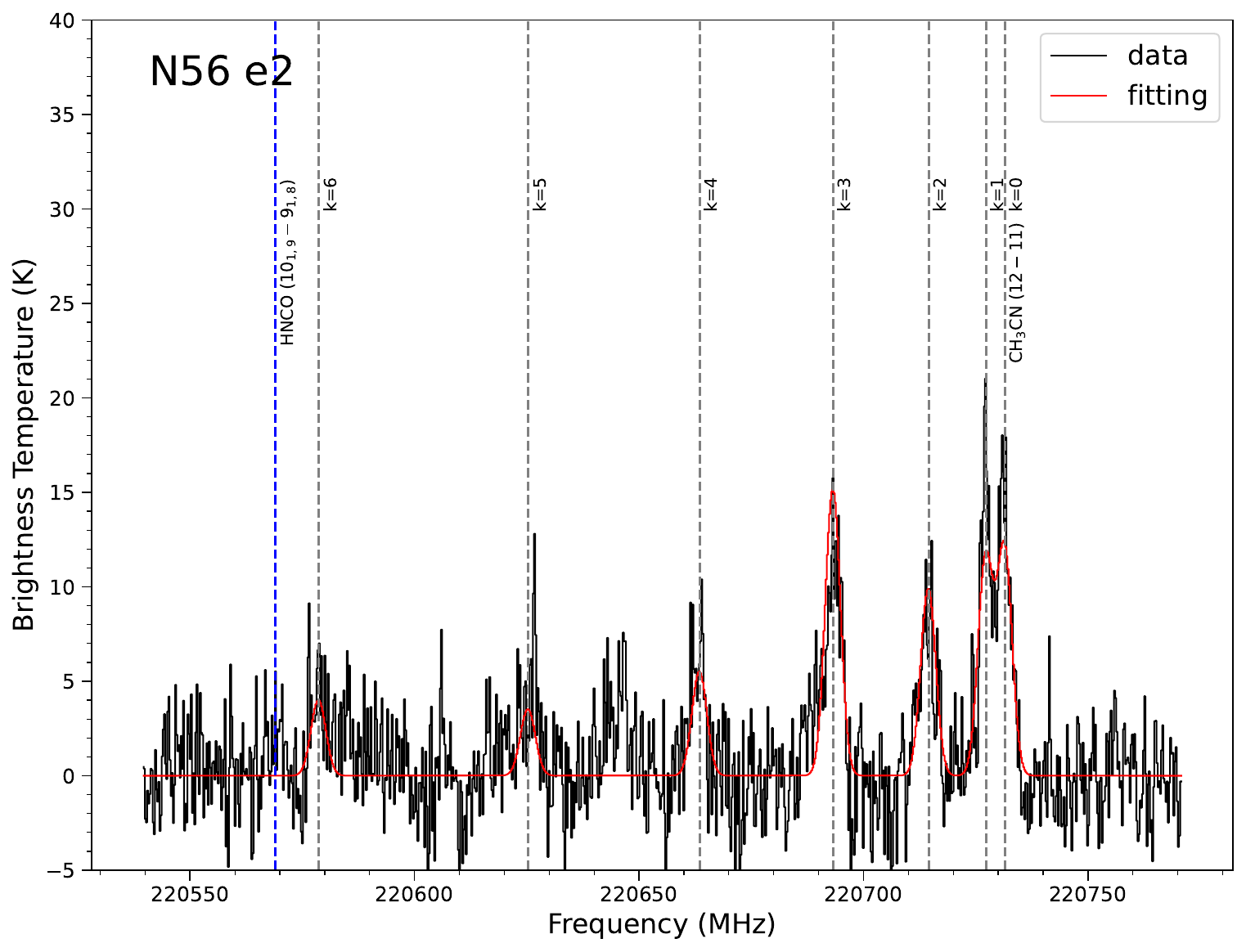}
    \includegraphics[width=0.45\textwidth]{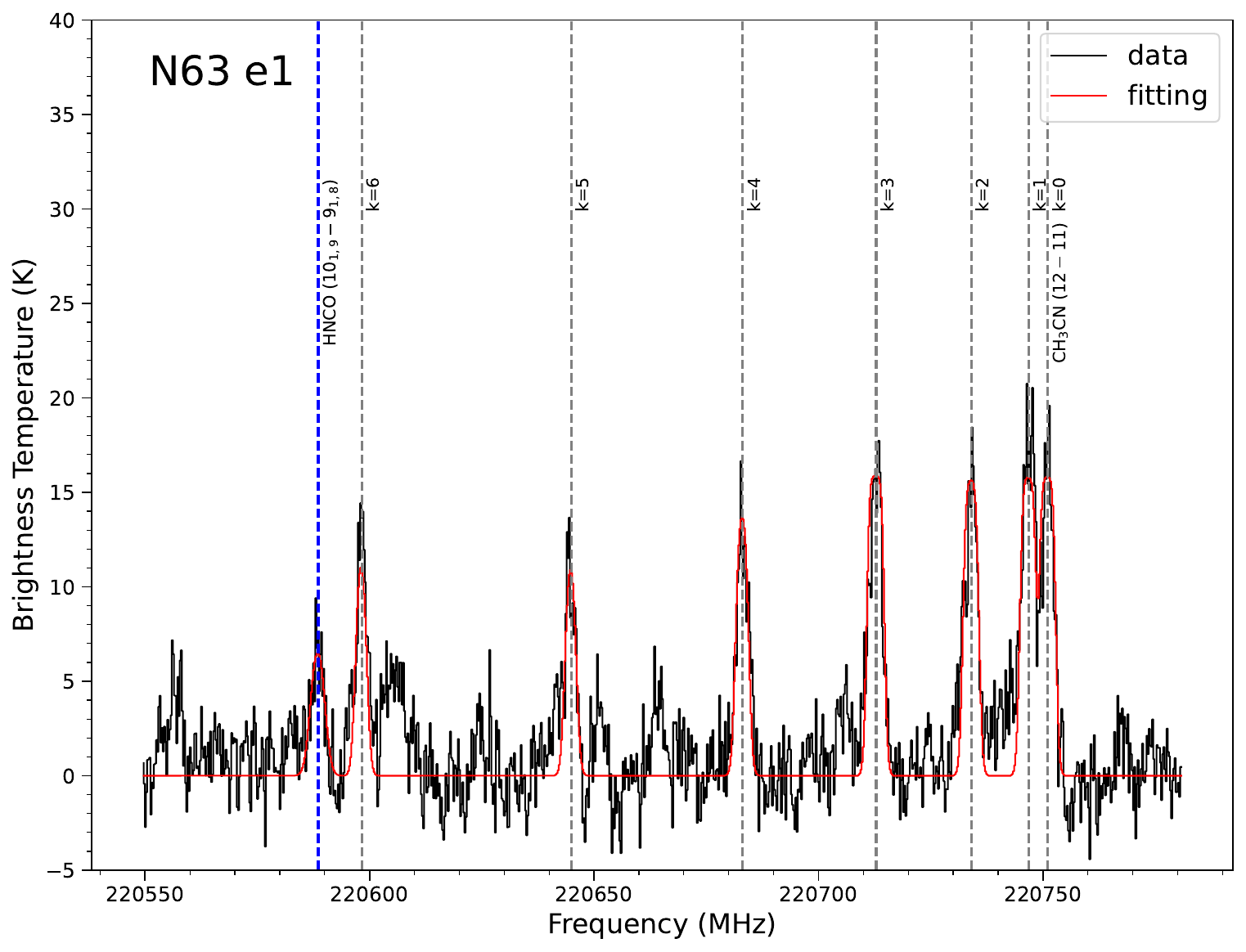}\\
    \includegraphics[width=0.45\textwidth]{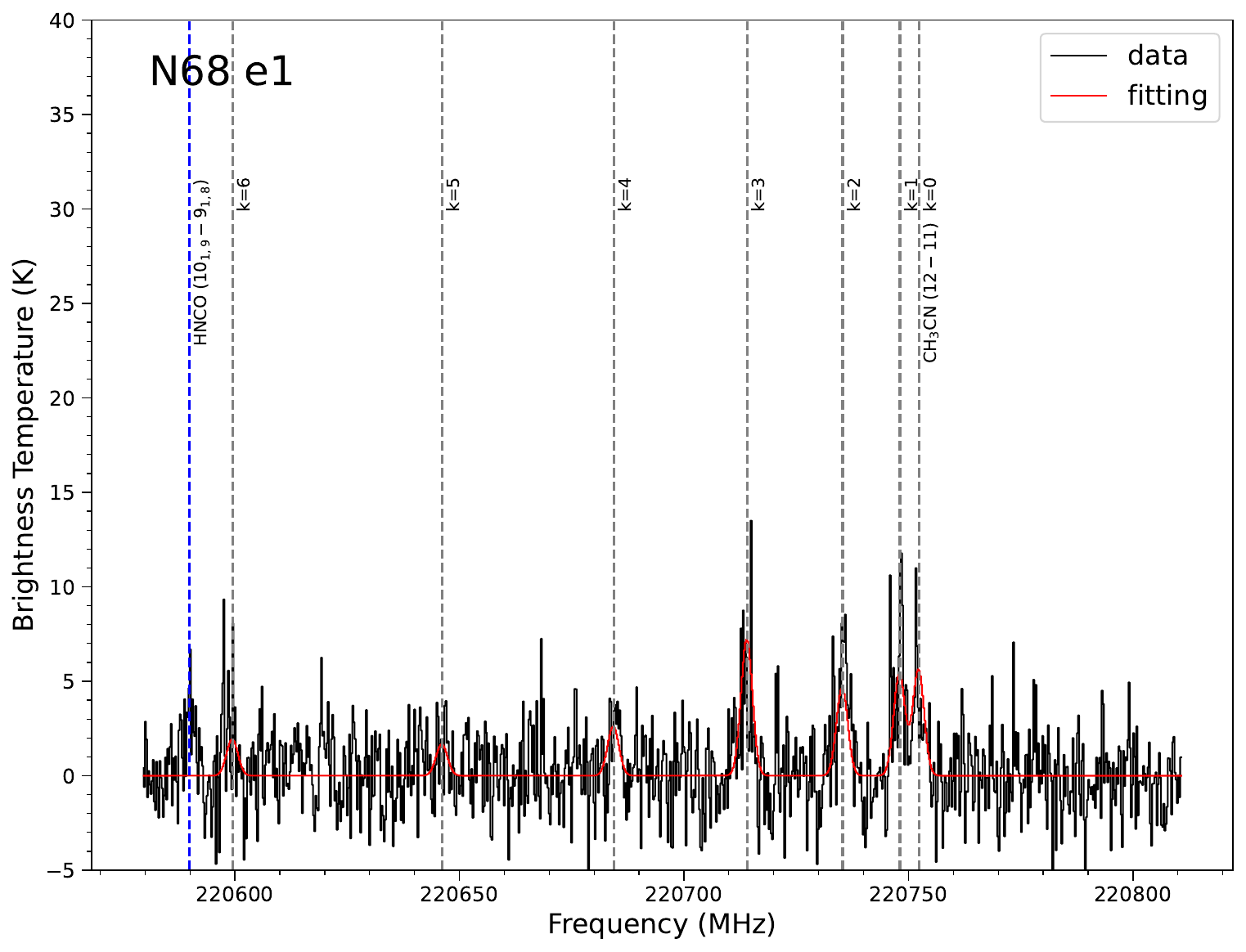}
    \includegraphics[width=0.45\textwidth]{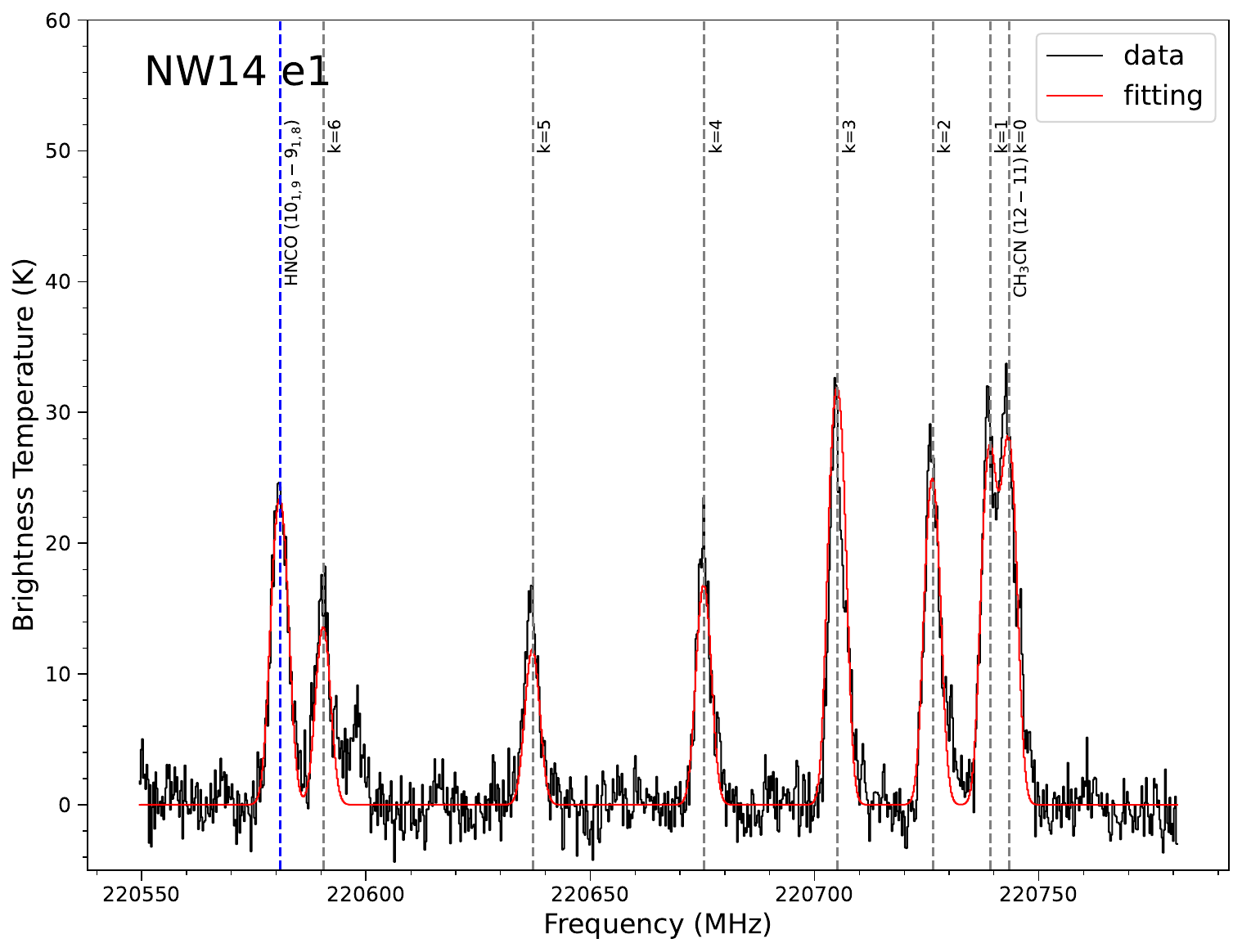}
    \caption{Same as Fig. \ref{fig:ch3cn_fitting_example}, but for each fragment detected in the $\mathrm{CH_3CN}~(12_K-11_K)~K=0-6$ line.}
    \label{fig:ch3cn_fitting_all}
\end{figure*}
\end{appendix}



%
%
\end{document}